\documentclass[12pt]{article}
\usepackage{a4,graphicx,amssymb}
\usepackage[small,bf]{caption}

\newcommand{\sixth}{\mbox{\small $\frac{1}{6}$}}         % 1/6
\newcommand{\half}{\mbox{\small $\frac{1}{2}$}}          % 1/2
\newcommand{\quart}{\mbox{\small $\frac{1}{4}$}}         % 1/4
\newcommand{\third}{\mbox{\small $\frac{1}{3}$}}         % 1/3
\newcommand{\twothird}{\mbox{\small $\frac{2}{3}$}}      % 2/3
\newcommand{\eightthird}{\mbox{\small $\frac{8}{3}$}}    % 8/3
\newcommand{\threequart}{\mbox{\small $\frac{3}{4}$}}    % 3/4
\newcommand{\qed}{\mbox{\tiny $Q\!E\!D$}}                % QED
\def\lsim{\mathrel{\rlap{\lower4pt\hbox{\hskip1pt$\sim$}}
    \raise1pt\hbox{$<$}}}                % less than or approx. symbol
\def\gsim{\mathrel{\rlap{\lower4pt\hbox{\hskip1pt$\sim$}}
    \raise1pt\hbox{$>$}}}                % greater than or approx. symbol

\begin{document}

\title{
\vspace{-3.25cm}
\flushright{\small ADP-12-26/T793} \\
\vspace{-0.35cm}
{\small DESY 12-094} \\
\vspace{-0.35cm}
{\small Edinburgh 2012/07} \\
\vspace{-0.35cm}
{\small Liverpool LTH 946} \\
\vspace{-0.35cm}
%{\small \today}  \\
%{\small June 14, 2012} \\
%{\small November 12, 2012} \\
{\small December 20, 2012} \\
\vspace{0.5cm}
\centering{\Large \bf Isospin breaking in octet baryon mass splittings}}

\author{\large
         R. Horsley$^a$, J. Najjar$^b$, \\
         Y. Nakamura$^c$, D. Pleiter$^d$, P.~E.~L. Rakow$^e$, \\
         G. Schierholz$^f$ and J.~M. Zanotti$^g$ \\[1em]
         -- QCDSF-UKQCD Collaboration -- \\[1em]
        \small $^a$ School of Physics and Astronomy,
               University of Edinburgh, \\[-0.5em]
        \small Edinburgh EH9 3JZ, UK \\[0.25em]
        \small $^b$ Institut f\"ur Theoretische Physik,
               Universit\"at Regensburg, \\[-0.5em]
        \small 93040 Regensburg, Germany \\[0.25em]
        \small $^c$ RIKEN Advanced Institute for
               Computational Science, \\[-0.5em]
        \small Kobe, Hyogo 650-0047, Japan \\[0.25em]
        \small $^d$ J\"ulich Supercomputer Centre,
               Forschungszentrum J\"ulich, \\[-0.5em]
        \small 52425 J\"ulich, Germany \\[0.25em]
        \small $^e$ Theoretical Physics Division,
               Department of Mathematical Sciences, \\[-0.5em]
        \small University of Liverpool,
               Liverpool L69 3BX, UK \\[0.25em]
        \small $^f$ Deutsches Elektronen-Synchrotron DESY, \\[-0.5em]
        \small 22603 Hamburg, Germany \\[0.25em]
        \small $^g$ CSSM, School of Chemistry and Physics,
               University of Adelaide, \\[-0.5em]
        \small Adelaide SA 5005, Australia}

%\date{\today}
%\date{May ??, 2011}
\date{}

\maketitle

% ----------------------------------------------------------------------

%\clearpage

\begin{abstract}
Using an $SU(3)$ flavour symmetry breaking expansion in the quark
mass, we determine the QCD component of the nucleon, Sigma and Xi
mass splittings of the baryon octet due to up-down (and strange)
quark mass differences in terms of the kaon mass splitting.
Provided the average quark mass is kept constant, the expansion
coefficients in our procedure can be determined from computationally
cheaper simulations with mass degenerate sea quarks and partially
quenched valence quarks. Both the linear and quadratic terms
in the $SU(3)$ flavour symmetry breaking expansion are considered;
it is found that the quadratic terms only change the result
by a few percent, indicating that the expansion is highly convergent.
\end{abstract}

%\clearpage

%----------------------------------------------------------------------------

\section{Introduction} 

%----------------------------------------------------------------------------

The masses of the baryon octet are now very accurately known,
with results given in the Particle Data Group \cite{nakamura10b} as
\begin{eqnarray}
   \begin{array}{lcllcl}
      M_n^{\exp} & = & 0.939565346(23)\,\mbox{GeV}\,,  &
      M_p^{\exp} & = & 0.938272013(23)\,\mbox{GeV}\,,      \\
      M_{\Sigma^-}^{\exp}
          & = & 1.197449(30)\,\mbox{GeV}\,,   &      
      M_{\Sigma^+}^{\exp} 
          & = & 1.18937(7)\,\mbox{GeV}\,,          \\
      M_{\Xi^-}^{\exp}
          & = & 1.32171(7)\,\mbox{GeV}\,,     &
      M_{\Xi^0}^{\exp}
          & = & 1.31486(20)\,\mbox{GeV}\,,        \\
   \end{array}
\label{Noctet_phys_masses}
\end{eqnarray}
around the outer ring of the octet and
\begin{eqnarray}
   \begin{array}{lcllcl}
      M_{\Sigma^0}^{\exp}
          & = & 1.192642(24)\, \mbox{GeV}\,,  &
      M_{\Lambda}^{\exp}
          & = & 1.115683(6)\, \mbox{GeV}\,,        \\
   \end{array}
\end{eqnarray}
at the centre. Isospin breaking effects (i.e.\ $u$ -- $d$
quark mass differences and electro\-magnetic effects) are
responsible for the nucleon, $n$ -- $p$, Sigma,
$\Sigma^-$ -- $\Sigma^+$, and Xi, $\Xi^-$ -- $\Xi^0$, mass splittings
\begin{eqnarray}
   (M_n - M_p)^{\exp} &=& 1.293333(33)\,\mbox{MeV}\,,
                                                           \nonumber    \\
   (M_{\Sigma^-} - M_{\Sigma^+})^{\exp}
             &=& 8.079(76)\,\mbox{MeV}\,,
                                              \label{expt_mass_split}   \\
   (M_{\Xi^-} - M_{\Xi^0})^{\exp}
             &=& 6.85(21)\,\mbox{MeV} \,.
                                                           \nonumber
\end{eqnarray}
These are very small differences (particularly for the $n$ -- $p$ mass
splitting), ranging from about $0.15\%$ to $0.7\%$ of the baryon mass.
(We shall not be considering the $\Sigma^0$ -- $\Lambda$ mass splitting
here as these particles have the same quantum numbers and 
mix if isospin is violated.) In this article we shall be only looking
at the hadronic or QCD contribution to these mass splittings, i.e.\ we
are not going to consider electromagnetic effects. Both effects
are small perturbations and can simply be added together. In the
case of the $n - p$ splitting we can argue that the hadronic
effect is larger because the electromagnetic effect would tend to 
make the proton heavier than the neutron (as $u$ quarks repel
more than $d$ quarks) which is not the case in the real world.
There have been several previous lattice determinations of the QCD
contribution to these mass splittings, e.g.\ 
\cite{beane06a,blum10a,divitiis11a}, and also several lattice
computations of the electromagnetic contribution, e.g.\
\cite{blum10a,duncan96a,blum07a,basak08a,portelli12a,glaessle11a,aoki12a}.
Non-lattice determinations include \cite{walker-lourd12a}.

\begin{figure}[htb]
   \begin{center}
      \includegraphics[width=6.50cm]
         {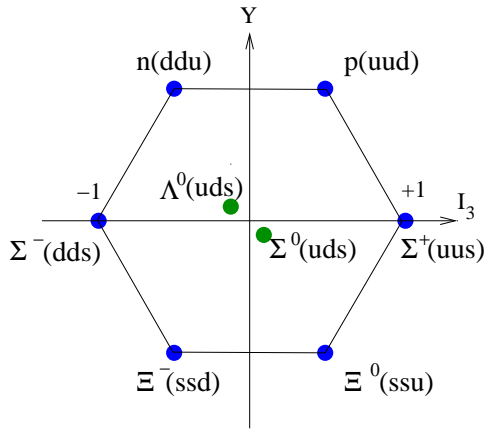}
   \end{center}
\caption{The lowest octet for the spin $\half$ baryons plotted
         in the $I_3$--$Y$ plane.}
\label{baryon_j=half_octet}
\end{figure}

In Fig.~\ref{baryon_j=half_octet} we sketch the lowest octet
for the spin $\half$ baryons plotted in the $I_3$--$Y$ plane.
The particles on the outer ring, namely the $n(ddu)$,
$p(uud)$, $\Sigma^+(dds)$, $\Sigma^-(uus)$ and $\Xi^-(ssd)$,
$\Xi^0(ssu)$ all consist of combinations of $aab$ quarks
(where we shall use the notation of denoting a quark, $q$,
by $a$, $b$, $\ldots$ which can be the up $u$, down $d$ or strange
$s$ quark). In this notation $a$ are the flavour doubly represented
quarks, while $b$ is the flavour singly represented quark.
At the centre of the octet we have two states, $\Lambda(uds)$ and
$\Sigma^0(uds)$, with the same quark content -- $u$, $d$ and $s$, but
different wavefunctions.

In \cite{bietenholz10a,bietenholz11a} we described a method for
extrapolating from the $SU(3)$ flavour symmetric point
(where we have three mass-degenerate quarks) to the physical point,
keeping the average of the quark masses constant.
The form of the $SU(3)$ flavour symmetry breaking expansion was developed
both for non-degenerate $u$, $d$ quark masses and for degenerate
$u$, $d$ quark masses. In \cite{bietenholz11a} numerical simulations
were performed for $2+1$ flavours, i.e.\ with two degenerate light
quark masses. Thus, effectively the mass `average' of the $n$, $p$
baryon, and the $\Sigma^+$, $\Sigma^-$ and $\Xi^-$, $\Xi^0$ baryons
were considered. However, as the coefficients of the quark mass
flavour symmetry are just functions of the average quark mass,
the expansion coefficients do not change from using non-degenerate
to $2$ or $3$ mass degenerate quarks provided that the average quark
mass is kept constant \cite{bietenholz11a}. This gives us the
opportunity to investigate isospin splittings, i.e.\ when the $u$
quark mass is different to the $d$ quark mass, using only results
from $2+1$ or $3$ flavour simulations.

As the baryon mass differences (e.g.\ $n$ -- $p$) depend on the
$u$ -- $d$ mass difference and are thus small, we find that
it is sufficient to consider the $SU(3)$ flavour symmetry
breaking expansion in the quark mass including both linear
terms (leading order, LO) and quadratic terms (next to
leading order, NLO). These LO and NLO terms were given in
\cite{bietenholz11a}.

We saw little curvature in hadron masses in the quark mass range
considered in \cite{bietenholz11a} (see also 
section~\ref{comparison_fan}), namely, from the
three degenerate flavour pion mass at $\sim 411\,\mbox{MeV}$
to the physical pion mass $\sim 140\,\mbox{MeV}$, so we conclude
that a much larger quark mass range is needed to reliably 
determine curvature. We achieve this larger range by extending
the numerical results to `{\it partially quenched}' or PQ quark
masses (where the valence quark masses do not have to be
the same as the sea or dynamical quark masses) with a spread of
quark masses from about one third of the strange quark mass
up to the charmed quark mass. We also consider part of the
next to next to leading order (NNLO) (cubic terms) in the $SU(3)$
flavour symmetry expansion.
(The NNLO terms were also indicated in \cite{bietenholz11a};
we have now completed this computation, \cite{QCDSF12a}.)
Thus we can consider a large quark mass range to be able to
determine the NLO or quadratic terms more accurately, using
the NNLO terms as a `control'.

Chiral perturbation theory looks at the breaking of the
chiral $SU(3)$ group by the quark masses - its expansion parameter
is the quark mass itself, or equivalently, the masses of the
pseudoscalar mesons. We are following an older strand, going back
to Gell-Mann and Okubo \cite{gell-mann62a,okubo62a} of looking
at the breaking of the non-chiral $SU(3)$ symmetry, by quark mass
differences. In chiral perturbation theory the expansion is about
the point where all three quarks are massless; here we expand about
a point where the three quarks have equal (non-zero) masses each
about a third of the physical strange quark mass.

As well as the baryon $SU(3)$ octet flavour expansion, we will also
need the values of the quark mass corresponding to the physical
point. We can achieve this by considering the equivalent
$SU(3)$ flavour expansion, but now for the pseudoscalar meson
octet. The same procedure as for the baryon octet is required:
first the $SU(3)$ flavour expansion coefficients must be
determined and then the experimental values of the masses
of the $K^0$, $K^+$ and  $\pi^+$ mesons can be used to
determine the required physical quark mass point.
These can then be used, together with the $SU(3)$ baryon octet
flavour expansion, to determine the mass splittings
for the baryon octet.

We shall find that the LO term is dominant (both for the
baryon and pseudoscalar octets) and so the NLO (and NNLO 
corrections) may be taken as an indication that our $SU(3)$
flavour symmetry breaking expansion appears to be a highly
convergent series. (This point is further discussed in
Appendix~\ref{systematic}.)

%----------------------------------------------------------------------------

\section{Octet Baryons}
\label{octet_baryons}

%----------------------------------------------------------------------------

Before discussing partial quenching, we first consider the 
case where the valence quark masses are the same as the sea
quark masses, the so-called `{\it unitary line}'. The
$SU(3)$ flavour symmetry breaking expansion,
\cite{bietenholz11a}, for all of the outer ring octet
baryons consisting of a pair of identical flavour quarks and
a third, different quark can be compactly written up to NLO as
\begin{eqnarray}
   M^2(aab)
      &=& M_0^2 + A_1 (2 \delta m_a + \delta m_b) 
               + A_2 (\delta m_b - \delta m_a)
                                                           \nonumber    \\
      & & \phantom{M_0^2} 
              + B_0 \sixth (\delta m_u^2 + \delta m_d^2 + \delta m_s^2 )
                                                 \label{N_1+1+1_exp}    \\
      & & \phantom{M_0^2} 
               + B_1 (2\delta m_a^2 + \delta m_b^2) 
               + B_2 (\delta m_b^2 - \delta m_a^2)
               + B_3 (\delta m_b - \delta m_a)^2 \,,
                                                           \nonumber 
\end{eqnarray}
with quarks $q = a$, $b$, $\ldots$ from $(u, d, s)$, where
\begin{eqnarray}
   \delta m_q  = m_q - \overline{m} \,, \quad
   \overline{m} = \third(m_u + m_d + m_s) \,.
\label{delta_mq_def}
\end{eqnarray}
We shall consider the $SU(3)$ flavour symmetry breaking
expansion of $M^2(aab)$ \cite{gasser82a}, rather than
$M(aab)$. Of course from the viewpoint of the $SU(3)$
flavour symmetry breaking expansion any function $f(M)$
could be considered. For fitting over a small quark mass range,
a linear function is sufficient; for the large quark mass
range considered in section~\ref{N_num_res} a better fit
to the numerical data was found using $M^2(aab)$
rather than $M(aab)$.

Note that $M_0^2$, $A_1$, $A_2$, $B_0$, $\ldots, B_3$
all depend on the average quark mass $\overline{m}$, which
will be held constant in the following simulations.
Keeping $\overline{m}$ constant reduces the number of
coefficients that must be determined (and indeed makes
the computation tractable).

From Fig.~\ref{baryon_j=half_octet} we see that as there
are six different masses on the baryon outer ring
but just two linear parameters in LO, the fits are
highly constrained. At the next order, NLO,
in eq.~(\ref{N_1+1+1_exp}) we are allowed four
coefficients for the quadratic terms.

We have in addition the trivial constraint
\begin{eqnarray}
   \delta m_u + \delta m_d + \delta m_s = 0 \,,
\label{sum_deltam}
\end{eqnarray}
so we can eliminate one of these quantities if we wish to.

Thus, to determine the octet baryon masses, we first have to
determine the expansion coefficients and second
we need to know the physical quark masses. In the following
we shall denote the physical point by a star ${}^*$.
We thus have two distinct computations. As we shall see,
the determination of the coefficients is helped by PQ simulations,
while $\delta m_q^*$ can be found by considering equivalent
expansions for the pseudoscalar meson octet.

A further problem is that the scale must be determined.
As discussed in \cite{bietenholz11a}, flavour blind (or
gluonic) quantities are suitable. We denote these
generically by $X$. One useful type of flavour blind quantity
can be considered as the `centre of mass' of the multiplet.
Thus for the baryon octet, one possibility is%
\footnote{Another independent possibility would be
$X_\Lambda^2 = \half( M_{\Lambda}^2 + M_{\Sigma^0}^2)$.}
\begin{eqnarray}
   X_N^2 = \sixth( M_p^2 + M_n^2 + M_{\Sigma^+}^2 +  M_{\Sigma^-}^2
                          + M_{\Xi^0}^2 + M_{\Xi^-}^2 ) \,.
\end{eqnarray}
At the physical point, from eq.~(\ref{Noctet_phys_masses}), this
gives
\begin{eqnarray}
   X_N^{\exp} = 1.1610\,\mbox{GeV} \,.
\label{XN_phys}
\end{eqnarray}
In general for the $SU(3)$ flavour breaking symmetry expansion
we have from eq.~(\ref{N_1+1+1_exp}),
\begin{eqnarray}
   X_N^2 &=& M_0^2 + (\sixth B_0+B_1+B_3)
                     (\delta m_u^2 + \delta m_d^2 + \delta m_s^2)
                                                           \nonumber    \\
        &=& M_0^2 + O(\delta m_q^2) \,.
\label{XN_expansion}
\end{eqnarray}
Upon adding the masses the $A_2$ and $B_2$ terms vanish; while the
$A_1$ term vanishes upon using the constraint equation,
eq~(\ref{sum_deltam}), and thus this leads to the vanishing of the
linear term in the quark mass. (This is indeed true for all flavour
blind quantities.)

Scale independent quantities can now be constructed by considering
the ratio $M^2(aab)/X_N^2$. Expanding to NLO order in the quark mass,
eq.~(\ref{N_1+1+1_exp}) retains the same pattern and becomes
\begin{eqnarray}
   \tilde{M}^2(aab)
      &=& 1 + \tilde{A}_1 (2 \delta m_a + \delta m_b) 
                  + \tilde{A}_2 (\delta m_b - \delta m_a)
                                                           \nonumber    \\
      & & \phantom{1} -(\tilde{B}_1+\tilde{B}_3)
                         (\delta m_u^2 + \delta m_d^2 + \delta m_s^2 )
                                            \label{Ntwid2_1+1+1_exp}    \\
      & & \phantom{1} + \tilde{B}_1 (2\delta m_a^2 + \delta m_b^2) 
              + \tilde{B}_2 (\delta m_b^2 - \delta m_a^2)
              + \tilde{B}_3 (\delta m_b - \delta m_a)^2 \,,
                                                           \nonumber
\end{eqnarray}
where a $\,\tilde{}\,$ on a hadron mass means that it has been divided
by $X_N$, so  $\tilde{M} = M/X_N$ while a $\,\tilde{}\,$ on the
expansion coefficients means that they have been divided by $M_0^2$,
for example $\tilde{A}_1 \equiv A_1/M_0^2$. From eq.~(\ref{XN_expansion})
we see that we have effectively replaced the $\sixth\tilde{B}_0$ term
by $-(\tilde{B}_1 + \tilde{B}_3)$. (However, this will not be important
in the case discussed here; as for mass differences, these terms
cancel again.)

Alternatively we can re-write eq.~(\ref{Ntwid2_1+1+1_exp}) as
\begin{eqnarray}
   \tilde{M}(aab)
      &=& 1 + \tilde{A}_1^\prime(2\delta m_a + \delta m_b) 
                    + \tilde{A}_2^\prime(\delta m_b - \delta m_a)
                                                         \nonumber   \\
      & &   - \half(\tilde{B}_1+\tilde{B}_3)
                          (\delta m_u^2 + \delta m_d^2 + \delta m_s^2)
                                           \label{Ntwid_1+1+1_exp}   \\
      & &   + \tilde{B}_1^\prime(2\delta m_a^2 + \delta m_b^2)
            + \tilde{B}_2^\prime(\delta m_b^2 - \delta m_a^2) 
            + \tilde{B}_3^\prime(\delta m_b - \delta m_a)^2  \,,
                                                         \nonumber
\end{eqnarray}
(essentially the equivalent $SU(3)$ flavour symmetry breaking expansion
for $M$ rather than $M^2$) with
\begin{eqnarray}
   \tilde{A}_1^\prime &=& \half \tilde{A}_1 \,,
                                                         \nonumber   \\
   \tilde{A}_2^\prime &=& \half \tilde{A}_2 \,,
                                                         \nonumber   \\
   \tilde{B}_1^\prime &=& \half(\tilde{B}_1-\threequart\tilde{A}_1^2) \,,
                                                         \nonumber   \\
   \tilde{B}_2^\prime &=& \half(\tilde{B}_2-\threequart\tilde{A}_1\tilde{A}_2)
                                          \,,
                                                         \nonumber   \\
   \tilde{B}_3^\prime &=& \half(\tilde{B}_3
              +\quart(2\tilde{A}_1-\tilde{A}_2)(\tilde{A}_1+\tilde{A}_2)) \,.
\label{sqrt_rels}
\end{eqnarray}

Although eq.~(\ref{Ntwid_1+1+1_exp}) looks complicated, we shall
only be interested in mass differences, which simplify
the expressions. Writing the flavour expansions as a Taylor
series in $\delta m_d \pm \delta m_u$ we find to NLO
\begin{eqnarray}
   \tilde{M}_n - \tilde{M}_p 
      &=& \tilde{M}(ddu) - \tilde{M}(uud)
                                                    \label{Nucsplit}  \\
      &=& (\delta m_d - \delta m_u)
          \left[ \tilde{A}_1^\prime -2\tilde{A}_2^\prime 
                + (\tilde{B}_1^\prime-2\tilde{B}_2^\prime)
                          (\delta m_d + \delta m_u)
          \right] \,,
                                                           \nonumber
\end{eqnarray}
together with
\begin{eqnarray}
   \tilde{M}_{\Sigma^-} - \tilde{M}_{\Sigma^+} 
      &=& \tilde{M}(dds) - \tilde{M}(uus)
                                                  \label{Sigmasplit}   \\
      &=& (\delta m_d - \delta m_u)
          \left[ 2\tilde{A}_1^\prime - \tilde{A}_2^\prime 
                + (2\tilde{B}^\prime_1-\tilde{B}_2^\prime + 3\tilde{B}_3^\prime)
                          (\delta m_d + \delta m_u)
          \right] \,,
                                                           \nonumber
\end{eqnarray}
and
\begin{eqnarray}
   \tilde{M}_{\Xi^-} - \tilde{M}_{\Xi^0} 
      &=& \tilde{M}(ssd) - \tilde{M}(ssu)
                                                     \label{Xisplit}    \\
      &=& (\delta m_d - \delta m_u)
          \left[ \tilde{A}_1^\prime + \tilde{A}_2^\prime 
                + (\tilde{B}_1^\prime + \tilde{B}_2^\prime + 3\tilde{B}_3^\prime)
                          (\delta m_d + \delta m_u)
          \right] \,,
                                                           \nonumber
\end{eqnarray}
where the constraint in eq.~(\ref{sum_deltam}) has been used to eliminate
$\delta m_s$. These equations represent the different types of isospin
differences possible for the baryon octet and are valid over a range
of quark masses. We shall consider them from the flavour symmetric point
down to the physical point, determined by
$\delta m_d^* - \delta m_u^*$ and 
$\delta m_d^* + \delta m_u^* = -\delta m_s^*$, $\delta m_q^*$ being the
physical point (which has to be determined).

We have a check of these formulas, since the Coleman-Glashow relation
\cite{coleman61a} should hold at this order,
\begin{eqnarray}
    ( \tilde{M}_n - \tilde{M}_p) 
      - (\tilde{M}_{\Sigma^-} - \tilde{M}_{\Sigma^+}) 
      + ( \tilde{M}_{\Xi^-} - \tilde{M}_{\Xi^0}) = O(\delta m_q^3) \,,
\label{cole-glash}
\end{eqnarray}
which is indeed satisfied by eqs.~(\ref{Nucsplit})--(\ref{Xisplit}).

\begin{figure}[htb]
   \begin{center}
      \includegraphics[angle=270, width=12cm]{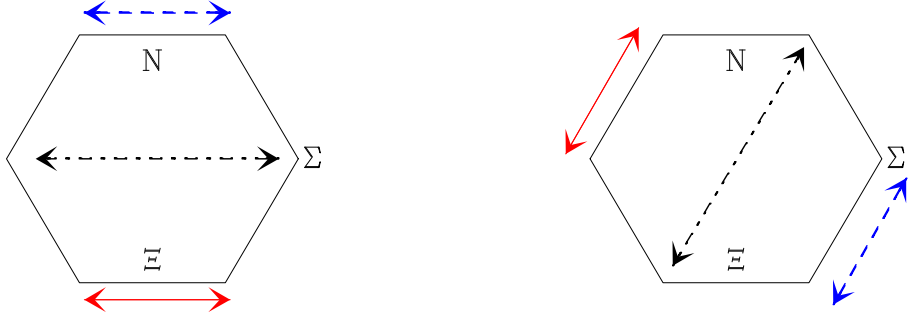}
    \end{center} 
\caption{Permutation group transformations link the  
         isospin violation caused by $m_d - m_u$ to the  
         $SU(3)$ violation caused by $m_s - m_u$, $m_s - m_d$.  
         The neutron-proton mass difference,
         $M_n - M_p$, dashed line, is mapped to
         $M_{\Xi^0} - M_{\Sigma^+}$; the $\Sigma$ splitting, 
         $M_{\Sigma^-} - M_{\Sigma^+}$, maps to
         $M_{\Xi^-} - M_p$ dot-dashed line; and
         $M_{\Xi^-} - M_{\Xi^0}$ is related to
         $M_{\Sigma^-} - M_n$, full line.}
\label{hexa}
\end{figure}

We can find relations between isospin violation caused by $m_d - m_u$
and to the $SU(3)$ violation caused by $m_s - m_u$, $m_s - m_d$ if we
make any $S_3$ transformation that changes the strange quark into
a light quark and vice versa (for example $u \to s \to d$ or
$d \leftrightarrow s$); see Fig.~\ref{hexa}.
($S_3$ is the symmetry group of the equilateral triangle and a
subgroup of $SU(3)$.) Applying the transformation $d \leftrightarrow s$
on both sides of eqs.~(\ref{Nucsplit})--(\ref{Xisplit}) we find to NLO
\begin{eqnarray}
   \tilde{M}_{\Xi^0} - \tilde{M}_{\Sigma^+} 
     &=&  (\delta m_s - \delta m_u ) \left[
             \tilde{A}_1^\prime - 2 \tilde{A}_2^\prime
               + (\tilde{B}_1^\prime - 2 \tilde{B}_2^\prime) 
                          (\delta m_s + \delta m_u)
                                     \right] \,,
                                                                        \\
   \tilde{M}_{\Xi^-} - \tilde{M}_p
     &=& (\delta m_s - \delta m_u) \left[
           2 \tilde{A}_1^\prime - \tilde{A}_2^\prime
             + (2 \tilde{B}_1^\prime - \tilde{B}_2^\prime + 3\tilde{B}_3^\prime) 
                         (\delta m_s + \delta m_u)
                                  \right] \,,
                                                           \nonumber    \\
   \tilde{M}_{\Sigma^-} - \tilde{M}_n 
     &=&  (\delta m_s - \delta m_u ) \left[
           \tilde{A}_1^\prime + \tilde{A}_2^\prime
               + (\tilde{B}_1^\prime + \tilde{B}_2^\prime + 3\tilde{B}_3^\prime)
                         (\delta m_s + \delta m_u)
                                  \right] \,.
                                                           \nonumber
 \end{eqnarray}

%----------------------------------------------------------------------------

\section{Determining the expansion coefficients}
\label{baryon_expan_coeffs}

%----------------------------------------------------------------------------

We now first find the $\tilde{A}_1$, $\tilde{A}_2$
and $\tilde{B}_1$, $\ldots$, $\tilde{B}_3$ coefficients.

%----------------------------------------------------------------------------

\subsection{Partially quenched octet baryons}
\label{pq_octet_baryons}

%----------------------------------------------------------------------------

Let us now generalise the previous results to the case when the
valence quarks do not have to have the same mass as the sea quarks
(i.e.\ we leave the unitary line). We have also generalised
eq.~(65) of \cite{bietenholz11a} from NLO to NNLO \cite{QCDSF12a}.
We find
\begin{eqnarray}
   M^2(aab) 
      &=& M_0^2 + A_1(2\delta\mu_a+\delta\mu_b) + A_2(\delta\mu_b-\delta\mu_a)
                                                         \nonumber  \\
      & & \phantom{M_0} 
              + \sixth B_0(\delta m_u^2 + \delta m_d^2 + \delta m_s^2)
                                                         \nonumber  \\
      & & \phantom{M_0} + B_1(2\delta\mu_a^2+\delta\mu_b^2)
                        + B_2(\delta\mu_b^2-\delta\mu_a^2) 
                        + B_3(\delta\mu_b-\delta\mu_a)^2
                                                         \nonumber   \\
      & & \phantom{M_0}
          + C_0\delta m_u\delta m_d\delta m_s
                                                         \nonumber   \\
      & & \phantom{M_0}
          + \left[ C_1(2\delta\mu_a + \delta\mu_b) 
                      + C_2(\delta\mu_b - \delta\mu_a)
                \right](\delta m_u^2 + \delta m_d^2 + \delta m_s^2)
                                                         \nonumber   \\
      & & \phantom{M_0}
                 + C_3(\delta\mu_a + \delta\mu_b)^3
                 + C_4(\delta\mu_a + \delta\mu_b)^2(\delta\mu_a - \delta\mu_b)
                                                         \nonumber   \\
      & &  \phantom{M_0}
                 + C_5(\delta\mu_a + \delta\mu_b)(\delta\mu_a - \delta\mu_b)^2
                 + C_6(\delta\mu_a - \delta\mu_b)^3 \,,
\label{N_1+1+1_pq}
\end{eqnarray}
where $a$, $b$, $\ldots$ now denote three valence quarks of arbitrary mass,
and we have defined
\begin{eqnarray}
   \delta\mu_q &=& \mu_q - \overline{m} \qquad q \in \{a, b, \ldots\} \,,
\label{delta_mu_bare}
\end{eqnarray}
where $\mu_q$ is the valence quark mass. In distinction to the
sea quarks, there is no restriction of the form of eq.~(\ref{sum_deltam})
on the values of valence quark masses. The numerical values of the
$M_0^2$, $A_1$, $A_2$, $B_0$, $B_1$, $\ldots$, $B_3$ and 
$C_0$, $\ldots$, $C_6$ coefficients are the same for PQ as for the
unitary case.

While we see that this is a relatively straightforward generalisation
of eq.~(\ref{N_1+1+1_exp}), we note that the term proportional
to $B_0$ remains unchanged. In addition the $C_0$ term also depends
entirely on sea terms, while the $C_1$ and $C_2$ terms are a mixture
of sea and valence terms. Thus if we wish to determine these
coefficients we must vary the sea quark masses; to determine the
other coefficients it is sufficient to vary the valence quark
masses alone, while keeping the sea quark masses constant.
So this gives the possibility of extending the (computationally
expensive) sea quark mass simulations with (computationally cheaper)
valence quark mass simulations to determine most of the coefficients.
If we work on a single sea background, then the $C_0$ 
term can be absorbed into the $M_0^2$ term, while the $C_1$ and $C_2$
terms can be absorbed into the $A_1$ and $A_2$ terms. If we
vary the sea quark masses this allows a determination of these
coefficients. However due to the constraint
$\overline{m} =\mbox{const.}$, or equivalently
eq.~(\ref{sum_deltam}), $\delta m_q$ cannot vary much and
we know from \cite{bietenholz11a} that in this range the
LO dominates, so these coefficients are difficult
to determine and contribute just noise. So practically we shall
ignore these terms in fits. (Alternatively, the constraint
$\overline{m} =\mbox{const.}$ could be relaxed, but then
we have additional expansion coefficients, which we wish to avoid.)
Thus in this article we regard the NNLO terms as `control'
on the LO and NLO terms.

As discussed in section~\ref{octet_baryons}, we can consider
scale independent quantities. Thus in analogy to
eq.~(\ref{Ntwid_1+1+1_exp}) we have
\begin{eqnarray}
   \tilde{M}^2(aab)
      &=& 1 + \tilde{A}_1 (2 \delta\mu_a + \delta\mu_b) 
                  + \tilde{A}_2 (\delta\mu_b - \delta\mu_a)
                                                           \nonumber    \\
      & & \phantom{1} - (\tilde{B}_1+\tilde{B}_3)
                         (\delta m_u^2 + \delta m_d^2 + \delta m_s^2 )
                                                           \nonumber    \\
      & & \phantom{1} + \tilde{B}_1 (2\delta\mu_a^2 + \delta\mu_b^2) 
              + \tilde{B}_2 (\delta\mu_b^2 - \delta\mu_a^2)
              + \tilde{B}_3 (\delta\mu_b - \delta\mu_a)^2
                                                         \nonumber   \\
      & & \phantom{1} 
             + (\tilde{C}_3-3\tilde{C}_5)\delta m_u\delta m_d\delta m_s
                                                         \nonumber   \\
      & & \phantom{1} + \left[ \tilde{C}_1(2\delta\mu_a + \delta\mu_b) 
                      + \tilde{C}_2(\delta\mu_b - \delta\mu_a)
                \right](\delta m_u^2 + \delta m_d^2 + \delta m_s^2)
                                                         \nonumber   \\
      & & \phantom{1} 
                 + \tilde{C}_3(\delta\mu_a + \delta\mu_b)^3
                 + \tilde{C}_4(\delta\mu_a + \delta\mu_b)^2
                                     (\delta\mu_a - \delta\mu_b)
                                                         \nonumber   \\
      & & \phantom{1}
                 + \tilde{C}_5(\delta\mu_a + \delta\mu_b)
                                     (\delta\mu_a - \delta\mu_b)^2
                 + \tilde{C}_6(\delta\mu _a - \delta\mu_b)^3 \,,
\label{N_1+1+1_pq_expt}
\end{eqnarray}
(where $X_N$ always depends just on the sea quarks and is given
by the NNLO extension of eq.~(\ref{XN_expansion})).

Furthermore, these equations remain valid if two of the sea quarks
are degenerate in mass, i.e. $m_u = m_d \equiv m_l$, the crucial 
point being that $\overline{m}$ must remain constant (as all the
coefficients are functions of $\overline{m}$). This means that
from dynamical $2+1$ flavour simulations we can determine the
$u$--$d$ mass splittings. The only change to eq.~(\ref{N_1+1+1_pq_expt})
when $m_u = m_d$ is that some terms become slightly simpler,
\begin{eqnarray}
   \delta m_u^2 + \delta m_d^2 + \delta m_s^2
      \to 6 \delta m_l^2 \,,
   \qquad \delta m_u\delta m_d \delta m_s \to - 2\delta m_l^3 \,,
\label{B_0_term}
\end{eqnarray}
where we have used the constraint equation, (\ref{sum_deltam}), which
now becomes
\begin{eqnarray}
   \delta m_s = -2\delta m_l \,.
\end{eqnarray}
This gives
\begin{eqnarray}
   \tilde{M}^2(aab)
      &=& 1 + \tilde{A}_1 (2 \delta\mu_a + \delta\mu_b) 
                  + \tilde{A}_2 (\delta\mu_b - \delta\mu_a)
                                                           \nonumber    \\
      & & \phantom{1} - 6(\tilde{B}_1+\tilde{B}_3)\delta m_l^2
                                                           \nonumber    \\
      & & \phantom{1} + \tilde{B}_1 (2\delta\mu_a^2 + \delta\mu_b^2) 
              + \tilde{B}_2 (\delta\mu_b^2 - \delta\mu_a^2)
              + \tilde{B}_3 (\delta\mu_b - \delta\mu_a)^2
                                                         \nonumber   \\
      & & \phantom{1} 
             -2 (\tilde{C}_3-3\tilde{C}_5)\delta m_l^3
                                                         \nonumber   \\
      & & \phantom{1} + 6\left[ \tilde{C}_1(2\delta\mu_a + \delta\mu_b) 
                      +  \tilde{C}_2(\delta\mu_b - \delta\mu_a)
                \right]\delta m_l^2
                                                         \nonumber   \\
      & & \phantom{1} 
                 + \tilde{C}_3(\delta\mu_a + \delta\mu_b)^3
                 + \tilde{C}_4(\delta\mu_a + \delta\mu_b)^2
                                     (\delta\mu_a - \delta\mu_b)
                                                         \nonumber   \\
      & & \phantom{1}
                 + \tilde{C}_5(\delta\mu_a + \delta\mu_b)
                                     (\delta\mu_a - \delta\mu_b)^2
                 + \tilde{C}_6(\delta\mu _a - \delta\mu_b)^3 \,,
\label{N_2+1_pq_expt}
\end{eqnarray}
with
\begin{eqnarray}
   X_N^2 = \third( M^2(lll) + M^2(lls) + M^2(ssl) ) \,.
\end{eqnarray}
(For a quark mass-degenerate $3$ flavour simulation
eq.~(\ref{N_2+1_pq_expt}) simplifies further as
$\delta m_l = 0 = \delta m_s$.)
In other words, using eq.~(\ref{N_2+1_pq_expt}) gives us all the
information we need to find the quark mass contribution relevant
for the $1+1+1$ case.

%----------------------------------------------------------------------------

\subsection{Numerical results}
\label{N_num_res}

%----------------------------------------------------------------------------

Simulations have been performed using $N_f = 2+1$ $O(a)$ improved
clover fermions \cite{cundy09a} at $\beta = 5.50$ and on
$32^3\times 64$ lattice sizes, as described in more detail
in \cite{bietenholz11a}. Errors given here are statistical
(using $\sim O(1500)$ configurations); possible systematic errors
are discussed in Appendix~\ref{systematic} and incorporated into
the final results in section~\ref{results}.

A particular starting value for the degenerate sea quark mass,
$m_0$, is chosen on the $SU(3)$ flavour symmetric line,
and the subsequent sea quark mass points $m_l$, $m_s$ have then
been arranged in the various simulations to have constant
$\overline{m}$ ($ = m_0$). This ensures that the expansion 
coefficients do not change. It was found in \cite{bietenholz11a}
that a linear fit provides a good description of the numerical data
over the relatively short distance from the symmetric point down
to the physical pion mass. This helped us in choosing the initial point
on the $SU(3)$ flavour symmetric line to give a path that hits
(or is very close to) the physical point.

In a little more detail, the bare quark masses in lattice units 
are defined as
\begin{eqnarray}
   m_q = {1 \over 2} 
            \left ({1\over \kappa_q} - {1\over \kappa_{0;c}} \right)
            \qquad \mbox{with} \quad q \in \{l, s, 0\} \,,
\label{kappa_bare}
\end{eqnarray}
(together with eq.~(\ref{delta_mq_def}) for $\delta m_q$)
with the index $q = 0$ denoting the common mass degenerate quarks
along the $SU(3)$ flavour symmetric line, and where vanishing of
the quark mass along this line determines $\kappa_{0;c}$. Keeping
$\overline{m} = \mbox{constant} \equiv  m_0$ gives
\begin{eqnarray}
   \kappa_s = { 1 \over { {3 \over \kappa_0} - {2 \over \kappa_l} } } \,.
\label{kappas_mbar_const}
\end{eqnarray}
So once we decide on a $\kappa_l$ this then determines $\kappa_s$.
Note that $\kappa_{0;c}$ drops out of eq.~(\ref{kappas_mbar_const}),
so we do not need its explicit value. The initial $SU(3)$ flavour
symmetric $\kappa_0$ value chosen here, namely $\kappa_0 = 0.12090$,
\cite{bietenholz11a} is very close to a point on the path that leads
to the physical point. The constancy of flavour singlet quantities
over the range from the $SU(3)$ flavour symmetric line down to
the physical point \cite{bietenholz11a}, leads directly from
$X_\pi$ to an estimate for the pion mass of $\sim 411\,\mbox{MeV}$
(i.e.\ eq.~(\ref{Xpi_phys})) and similarly from $X_N$ a value of
the lattice spacing of $a \sim 0.079\,\mbox{fm}$.

While, as discussed earlier, simulations between the $SU(3)$ 
flavour symmetric point and the physical point are in principle
enough to determine the linear and quadratic expansion coefficients,
in practice the range is not sufficiently large to reliably determine
the quadratic terms. In order to determine the quadratic coefficients
more precisely, additional PQ simulations have been performed on the set
of gauge configurations that have all three sea quark masses equal,
i.e.\ at the $SU(3)$ flavour symmetric point $\kappa_0 = 0.12090$.
For these particular simulations $\delta m_l = 0 = \delta m_s$ automatically.
$\mu_q$ is defined identically to $m_q$, eq~(\ref{kappa_bare}),
by replacing $m_q \to \mu_q$ with $q \in \{ a, b, \ldots, \}$ together
with  eq.~(\ref{delta_mu_bare}) for $\delta \mu_q$ so that
\begin{eqnarray}
   \delta\mu_q = \mu_q - \overline{m}\,, \qquad
   \mu_q = {1 \over 2} 
            \left ({1\over \kappa_q} - {1\over \kappa_{0;c}} \right) \,.
\end{eqnarray}
We have chosen a wide range of PQ masses starting from a slightly 
heavier mass than $m_0$ (to avoid any possibility of so-called
`exceptional configurations') and reaching up to masses $\sim m_{charm}$.
The values are given in Appendix~\ref{pq_mass_tables} in
Table~\ref{pq_N_masses_I}. A direct fit is made to these PQ masses,
the unitary masses of \cite{bietenholz11a}, which all have the same
fixed $\overline{m}$, and additionally three PQ masses,
which we call $M_{N_s} \equiv M(sss)$ again with the same fixed
$\overline{m}$. (This is analogous to the pseudoscalar $\eta_s$
considered later.) These values are also given in
Appendix~\ref{pq_mass_tables} in Table~\ref{pq_Ns_masses}.
Using the fit function of eq.~(\ref{N_1+1+1_pq_expt}) and ignoring
the $\tilde{C}_1$ and $\tilde{C}_2$ terms as discussed in 
section~\ref{pq_octet_baryons} gives the results in
Table~\ref{Noctet_parms}, together with
%\mnote{120508\_1643}
\begin{table}[htb]
   \begin{center}
      \begin{tabular}{cc|ccc}
         $\tilde{A}_1$ & $\tilde{A}_2$ &
         $\tilde{B}_1$ & $\tilde{B}_2$ & $\tilde{B}_3$ \\
         \hline
         10.15(12) & 1.828(157) & 13.51(126) & -10.29(139) & -14.90(144) \\
      \end{tabular}
   \end{center}
\caption{Results for the baryon octet expansion coefficients.}
\label{Noctet_parms}
\end{table}
the NNLO coefficient values of $\tilde{C}_3 = -4.60(115)$,
$\tilde{C}_4 = -17.5(26)$, $\tilde{C}_5 = -1.88(310)$ and
$\tilde{C}_6 = -3.65(184)$. We note that in comparison to the
NLO coefficients, the NNLO coefficients are poorly determined.
The bootstrap (MINUIT) fit used gave $\chi^2/\mbox{dof} \sim 0.4$. 

\begin{figure}[htb]
   \begin{center}
      \includegraphics[width=9.00cm]
         {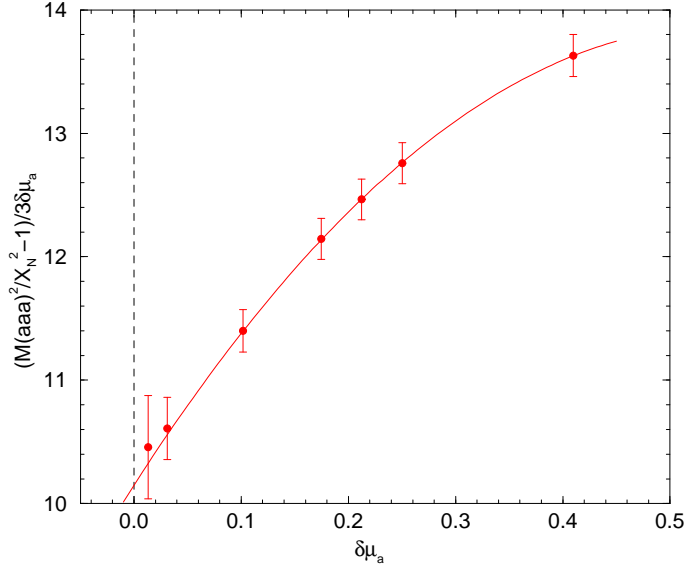}
   \end{center}
\caption{$(\tilde{M}^2(aaa)-1)/(3\delta\mu_a)$ against $\delta\mu_a$,
         together with eq.~(\protect\ref{Maaa}) using the $\tilde{A}_1$,
         $\tilde{B}_1$ fit values from Table~\protect\ref{Noctet_parms}
         and $\tilde{C}_3$ (given in the text).}
\label{dmua_mNaaaoXN}
\end{figure}

We now compare these results to a plot for illustration.
The simplest to consider is setting $\delta\mu_a = \delta\mu_b$,
i.e.\ degenerate valence quark masses. For simplicity, but
slightly inaccurately, we shall in the following simply say $b = a$.
(Of course we still need two different quark flavours.)
With $\delta m_l = 0$, eq.~(\ref{N_2+1_pq_expt}) then becomes
\begin{eqnarray}
   {\tilde{M}^2(aaa)-1 \over 3\delta\mu_a}
      &=& \tilde{A}_1
          + \tilde{B}_1\delta\mu_a + \eightthird\tilde{C}_3\delta\mu_a^2 \,.
\label{Maaa}
\end{eqnarray}
In Fig.~\ref{dmua_mNaaaoXN} we plot
$(\tilde{M}^2(aaa)-1)/(3\delta\mu_a)$ against $\delta\mu_a$ for the
PQ data (together with the cubic fit coefficients from
Table~\ref{Noctet_parms}) which due to the denominator is a sensitive plot.
There is good agreement. (We postpone the comparison to the unitary
data, `fan' plots, until section~\ref{comparison_fan}.)

%----------------------------------------------------------------------------

\section{Octet pseudoscalar mesons}

%----------------------------------------------------------------------------

Determining the octet pseudoscalar mass splittings (or more accurately
the splittings of the quadratic masses) will give
$\delta m_u^*$, $\delta m_d^*$ and $\delta m_s^*$ (the quark masses at the
physical point). This closely follows the baryon octet procedure,
we must again consider the analogous flavour symmetry expansion
for the pseudoscalar meson octet together
with the known experimental masses of the pions and kaons.

\begin{figure}[htb]
   \begin{center}
      \includegraphics[width=6.50cm]
         {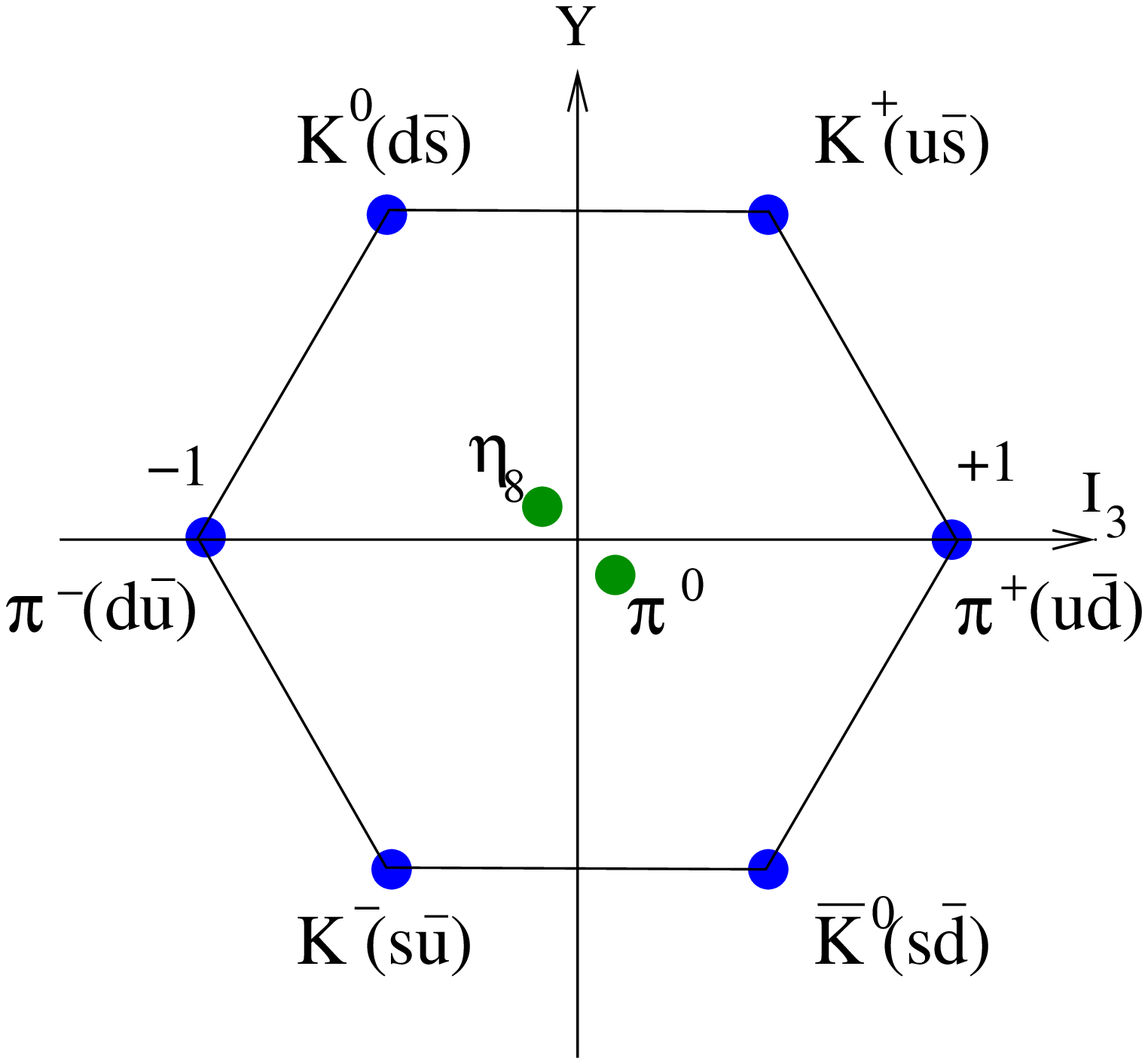}
   \end{center}
\caption{The lowest octet for the spin $0$ pseudoscalar mesons
         plotted in the $I_3$--$Y$ plane.}
\label{meson_j=0_octet}
\end{figure}

In Fig.~\ref{meson_j=0_octet} we sketch the lowest pseudoscalar octet
in the $I_3$--$Y$ plane. We have $K^0(d\overline{s})$, $K^+(u\overline{s})$,
$\pi^+(u\overline{d})$ together with $\overline{K}^0(s\overline{d})$,
$K^-(s\overline{u})$ and $\pi^-(d\overline{u})$ in the (outer)
ring of the octet. From charge conjugation or $C$ invariance
we further have $M_{\overline{K}^0} = M_{K^0}$, $M_{K^-} = M_{K^+}$,
and $M_{\pi^-} = M_{\pi^+}$ (which is in distinction to
the baryon octet which does not have this constraint).

% -------------------------------------------------------------------------

\subsection{PQ pseudoscalar meson flavour expansions}
\label{pq_ps_flavour_expansions}

% -------------------------------------------------------------------------

The corresponding formulas for the octet pseudoscalar mesons
are simpler than for the octet baryon, due to the constraints
imposed by $C$ invariance. The following $SU(3)$ flavour breaking
expansion formula is always valid for quarks $q = a$, $b$, $\ldots,$
in $(u, d, s, \ldots,)$
\begin{eqnarray}
   M^2(a\overline{b})
      &=& M^2_{0\pi} + \alpha(\delta m_a + \delta m_b)
                                            \label{M2_exp_unitary}  \\
      & & \phantom{M^2_{0\pi}}
             + \beta_0\sixth(\delta m_u^2 + \delta m_d^2 + \delta m_s^2)
             + \beta_1(\delta m_a^2 + \delta m_b^2)
             + \beta_2(\delta m_a - \delta m_b)^2 \,,
                                                         \nonumber
\end{eqnarray}
in quark masses up to the NLO as discussed in \cite{bietenholz11a}.
(Note that $M(a\overline{b}) = M(b\overline{a})$.)

Again combinations of masses can be chosen, so that
due to eq.~(\ref{sum_deltam}) the linear term in
eq.~(\ref{M2_exp_unitary}) vanishes, which is equivalent to 
averaging the outer ring of particles and finding the
`centre of mass' of the octet. In particular if we set
\begin{eqnarray}
   X_{\pi}^2 &=& \sixth( M_{K^+}^2 + M_{K^0}^2 + M_{\pi^+}^2 
                     + M_{\pi^-}^2 + M_{\overline{K}^0}^2 + M_{K^-}^2)
                                                           \nonumber \\
           &=& \third( M_{K^+}^2 + M_{K^0}^2 + M_{\pi^+}^2 ) \,,
\label{Xpi_def}
\end{eqnarray}
this gives
\begin{eqnarray}
   X_{\pi}^2 
      &=& M_{0\pi}^2 
          + \left( \sixth\beta_0 + \twothird\beta_1 + \beta_2 \right)
                  (\delta m_u^2 + \delta m_d^2 + \delta m_s^2 )
                                                           \nonumber \\
      &=& M_{0\pi}^2 + O(\delta m_q^2) \,.
\end{eqnarray}

In the partially quenched case, the above $SU(3)$ flavour expansion
can be generalised to
\begin{eqnarray}
   M^2(a\overline{b})
      &=& M^2_{0\pi} + \alpha(\delta\mu_a + \delta\mu_b)
                                                         \nonumber   \\
      & & \phantom{M^2_{0\pi}}
            + \beta_0\sixth(\delta m_u^2 + \delta m_d^2 + \delta m_s^2)
            + \beta_1(\delta\mu_a^2 + \delta\mu_b^2)
            + \beta_2(\delta\mu_a - \delta\mu_b)^2
                                                         \nonumber   \\
      & & \phantom{M^2_{0\pi}}
            + \gamma_0\delta m_u\delta m_d\delta m_s
            + \gamma_1(\delta\mu_a + \delta\mu_b)
                       (\delta m_u^2 + \delta m_d^2 + \delta m_s^2)
                                                         \nonumber   \\
      & & \phantom{M^2_{0\pi}}
           + \gamma_2(\delta\mu_a + \delta\mu_b)^3
           + \gamma_3(\delta\mu_a + \delta\mu_b)
                              (\delta\mu_a - \delta\mu_b)^2 \,,
\end{eqnarray}
where the NLO was also discussed in \cite{bietenholz11a} and again
we have extended the formula to the NNLO case \cite{QCDSF12a}.
This is again a general formula valid for possibly differing masses
of the sea, $m_q$, and valence quarks, $\mu_q$. The same notation
has been used as in sections \ref{octet_baryons} and
\ref{baryon_expan_coeffs}. (In particular remember that $\delta\mu_q$
is unconstrained in distinction to the sea quarks, eq.~(\ref{sum_deltam}).
The unitary line is recovered when $\mu_q \to m_q$.) Again as
with the PQ octet baryon case, eq.~(\ref{N_1+1+1_pq}), we see that
at the NNLO there is a term, the $\gamma_1$ term, which mixes the
sea and valence quarks; as discussed in section~\ref{pq_octet_baryons}
this again makes the numerical determination of these coefficients
difficult (we cannot vary the sea quark masses over a large enough range).
So in the same spirit as section~\ref{pq_octet_baryons} we ignore
this term and regard the NNLO terms as `control' terms.

If we wish to consider `physical ratios' then the masses can again
be conveniently normalised using $X_\pi^2$. Expanding to NNLO,
we find
\begin{eqnarray}
   \tilde{M}^2(a\overline{b})
      &=& 1 + \tilde{\alpha}(\delta\mu_a + \delta\mu_b)
                                                         \nonumber   \\
      & & \phantom{1}
            - (\twothird\tilde{\beta}_1 + \tilde{\beta}_2)
              (\delta m_u^2 + \delta m_d^2 + \delta m_s^2)
                 + \tilde{\beta}_1(\delta\mu_a^2 + \delta\mu_b^2)
                 + \tilde{\beta}_2(\delta\mu_a - \delta\mu_b)^2
                                                         \nonumber   \\
      & & \phantom{1}
            + 2(\tilde{\gamma}_2-3\tilde{\gamma}_3)
                             \delta m_u\delta m_d\delta m_s
            + \tilde{\gamma}_1(\delta\mu_a + \delta\mu_b)
                             (\delta m_u^2 + \delta m_d^2 + \delta m_s^2)
                                                         \nonumber   \\
      & & \phantom{1}
            + \tilde{\gamma}_2(\delta\mu_a + \delta\mu_b)^3
            + \tilde{\gamma}_3(\delta\mu_a + \delta\mu_b)
                              (\delta\mu_a - \delta\mu_b)^2 \,,
\label{pq_fit_meson_1+1+1}
\end{eqnarray}
where again a $\,\tilde{}\,$ on a hadron mass squared means that it
has been divided by $X_\pi^2$ (which only depends on the sea quarks)
while on an expansion coefficient it means that the coefficient
has been divided by $M_{0\pi}^2$ for example
$\tilde{\alpha} = \alpha / M_{0\pi}^2$.

Again for a $2+1$ flavour simulation (the case that will be considered
here) eq.~(\ref{pq_fit_meson_1+1+1}) slightly simplifies to become
\begin{eqnarray}
   \tilde{M}^2(a\overline{b})
      &=& 1 + \tilde{\alpha}(\delta\mu_a + \delta\mu_b)
                                                         \nonumber   \\
      & & \phantom{1}
            - 2(2\tilde{\beta}_1 + 3\tilde{\beta}_2)\delta m_l^2
                 + \tilde{\beta}_1(\delta\mu_a^2 + \delta\mu_b^2)
                 + \tilde{\beta}_2(\delta\mu_a - \delta\mu_b)^2
                                                         \nonumber   \\
      & & \phantom{1}
            - 4(\tilde{\gamma}_2-3\tilde{\gamma}_3)\delta m_l^3
            + 6\tilde{\gamma}_1(\delta\mu_a + \delta\mu_b)\delta m_l^2
                                                         \nonumber   \\
      & & \phantom{1}
            + \tilde{\gamma}_2(\delta\mu_a + \delta\mu_b)^3
            + \tilde{\gamma}_3(\delta\mu_a + \delta\mu_b)
                              (\delta\mu_a - \delta\mu_b)^2 \,,
\label{pq_fit_meson_2+1}
\end{eqnarray}
with the same coefficients, provided of course that $\overline{m}$
remains constant and where
\begin{eqnarray}
   X_\pi^2 = \third(2M^2(l\overline{s}) + M^2(l\overline{l}) ) \,.
\label{Xpi_lsdef}
\end{eqnarray}

% -------------------------------------------------------------------------

\subsection{Determination of the coefficients}

% -------------------------------------------------------------------------

As in section~\ref{baryon_expan_coeffs} we need to consider PQ
masses at the $SU(3)$ flavour symmetric point for the sea quark masses
(see Table~\ref{pq_ps_masses}) jointly with the $32^3\times 64$ lattice size
unitary results from Table~20 of \cite{bietenholz11a} all with the same
$\overline{m}$ constant value. From eq.~(\ref{pq_fit_meson_2+1})
we need to determine the constants
$\tilde{\alpha}$, $\tilde{\beta_1}$, $\tilde{\beta_2}$ and
$\tilde{\gamma}_2$, $\tilde{\gamma}_3$.

A $5$-parameter fit to the PQ and unitary data then yields the results of
Table~\ref{alpha_beta_results} together with the NNLO coefficient values of
%\mnote{120508\_1536}
\begin{table}[htb]
   \begin{center}
      \begin{tabular}{c|cc}
         $\tilde{\alpha}$ & $\tilde{\beta}_1$ & $\tilde{\beta}_2$ \\
         \hline
         41.17(8) & 76.50(148) & -45.81(189) \\
      \end{tabular}
   \end{center}
\caption{Results for the pseudoscalar octet expansion coefficients.}
\label{alpha_beta_results}
\end{table}
$\tilde{\gamma}_2 = -16.4(12)$ and $\tilde{\gamma}_3 = -20.3(39)$.
The bootstrap (MINUIT) fit used gave $\chi^2/\mbox{dof} \sim 1.7$. 

\begin{figure}[htbp]
   \begin{center}
      \includegraphics[width=9.00cm]
         {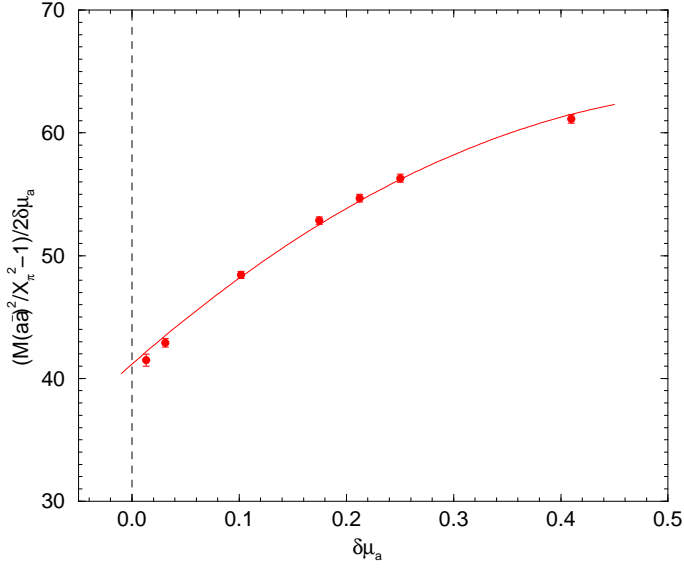}
   \end{center}
\caption{$(\tilde{M}^2(a\overline{a})-1)/(2\delta\mu_a)$ against
         $\delta\mu_a$. The full line is eq.~(\protect\ref{Maa}),
         using the $\tilde{\alpha}$, $\tilde{\beta}_1$ fit values
         from Table~\protect\ref{alpha_beta_results} and
         $\tilde{\gamma}_2$ (given in the text).}
\label{dmua_mpsaa2oXps2}
\end{figure}

As in section~\ref{N_num_res}, it is useful to compare these fit
results in a plot. In parallel to eq.~(\ref{Maaa}) let us again
consider the simple case of degenerate quark mass, i.e.\ where quark
and antiquark have the same mass, but different flavours, so they
cannot annihilate. We set $b = a$ and $\delta m_l = 0$ in
eq.~(\ref{pq_fit_meson_2+1}) and so consider
\begin{eqnarray}
    {\tilde{M}^2(a\overline{a}) - 1 \over 2\delta\mu_a}
       &=& \tilde{\alpha}
             + \tilde{\beta}_1\delta\mu_a
             + 4\tilde{\gamma}_2\delta\mu_a^2 \,.
\label{Maa}
\end{eqnarray}
In Fig.~\ref{dmua_mpsaa2oXps2} we plot
$(\tilde{M}^2(a\overline{a})-1)/(2\delta\mu_a)$ against $\delta\mu_a$
using the PQ data. This is compared with the cubic fit from eq.~(\ref{Maa})
and Table~\ref{alpha_beta_results}. There is good agreement.

% -------------------------------------------------------------------------

\subsection{Pseudoscalar meson isospin splittings}
\label{ps_meson_isobrk}

% -------------------------------------------------------------------------

Having determined $\tilde{\alpha}$, $\tilde{\beta}_1$, $\tilde{\beta}_2$
we can now find $\delta m_u$, $\delta m_d$, $\delta m_s$ given the
masses around the outer ring of the pseudoscalar octet. We postpone a
discussion of the numerical values until the next section and here just
derive the relevant formulas. As they are trivially valid over a range
of quark masses (and not just at the physical point) we give
these more general expressions here. 

From eqs.~(\ref{Nucsplit})--(\ref{Xisplit}) we see that we need
$\delta m_d - \delta m_u \equiv \delta m_-$ and
$\delta m_d + \delta m_u \equiv \delta m_+$.
So as in section~\ref{octet_baryons} we consider the mass difference
\begin{eqnarray}
   \tilde{M}_{K^0}^2 - \tilde{M}_{K^+}^2
     &=& (\delta m_d - \delta m_u)
         [\tilde{\alpha} 
           + (\tilde{\beta}_1+3\tilde{\beta}_2)(\delta m_d + \delta m_u)]
                                                           \nonumber \\
     &=&  \delta m_-[\tilde{\alpha} + 
                       (\tilde{\beta}_1+3\tilde{\beta}_2)\delta m_+] \,.
\label{K0-Kstar_phys}
\end{eqnarray}
Alternatively we can consider
\begin{eqnarray}
   \tilde{M}_{K^+}^2 - \tilde{M}_{\pi^+}^2
     &=& (\delta m_s - \delta m_d)
         [\tilde{\alpha} 
           + (\tilde{\beta}_1+3\tilde{\beta}_2)(\delta m_s + \delta m_d)]
                                                           \nonumber \\
     &=& -\half(\delta m_- + 3\delta m_+)
         [\tilde{\alpha} 
           + \half(\tilde{\beta}_1+3\tilde{\beta}_2)
                    (\delta m_- - \delta m_+)] \,,
\label{Kstar-pistar_phys}
\end{eqnarray}
and
\begin{eqnarray}
   \tilde{M}_{K^0}^2 - \tilde{M}_{\pi^+}^2
     &=& (\delta m_s - \delta m_u)
         [\tilde{\alpha}
           + (\tilde{\beta}_1+3\tilde{\beta}_2)(\delta m_s + \delta m_u)]
                                                           \nonumber \\
     &=& \half(\delta m_- - 3\delta m_+)
         [\tilde{\alpha} 
           - \half(\tilde{\beta}_1+3\tilde{\beta}_2)
                    (\delta m_- + \delta m_+)] \,,
\label{pistar-K0_phys}
\end{eqnarray}   
where we have used the constraint, eq.~(\ref{sum_deltam}),
to eliminate $\delta m_s$ in the second line of
eqs.~(\ref{Kstar-pistar_phys}) and (\ref{pistar-K0_phys})
to re-write the equations in terms of $\delta m_-$, $\delta m_+$.

Of course only two of the equations
(\ref{K0-Kstar_phys})--(\ref{pistar-K0_phys})
are independent. Choosing eqs.~(\ref{K0-Kstar_phys}) and
(\ref{Kstar-pistar_phys}), these quadratic equations can be solved
iteratively to give $\delta m_\mp$. We start
the iteration from the linear term alone or LO
in the quark mass. From eq.~(\ref{K0-Kstar_phys}) we see that is 
sufficient to determine $\delta m_+$ to LO (and $\delta m_-$
to NLO). Thus we find
\begin{eqnarray}
   \delta m_d - \delta m_u
      &=& {\tilde{M}_{K^0}^2 - \tilde{M}_{K^+}^2 \over 
                               \tilde{\alpha}} \,
          \left( 
           1 + {2(\tilde{\beta}_1 + 3\tilde{\beta}_2) \over 3\tilde{\alpha}^2}
               (\half(\tilde{M}_{K^0}^2+\tilde{M}_{K^+}^2)
                                     -\tilde{M}_{\pi^+}^2)
          \right) \,,
                                                           \nonumber \\
   \delta m_d + \delta m_u
      &=& - {2 \over 3\tilde{\alpha}}\,
               \left( \half(\tilde{M}_{K^0}^2+\tilde{M}_{K^+}^2)
                                     -\tilde{M}_{\pi^+}^2
               \right)[ 1  + \epsilon_{NLO} ] \,.
\label{iter_dmpm}
\end{eqnarray}
As $2(\tilde{\beta}_1 + 3\tilde{\beta}_2) / 3\tilde{\alpha}^2 \sim -0.02$
then as, in particular, at the physical point
$\half(\tilde{M}_{K^0}^{*\,2}+\tilde{M}_{K^+}^{*\,2}) -\tilde{M}_{\pi^+}^{*\,2}$
is $\sim 1$, we expect that the NLO term for $\delta m_-$
is small,%
\footnote{This is also true for $\delta m_+$. The NLO term is
\begin{eqnarray}
   \epsilon_{NLO}
      = - {\tilde{\beta}_1+3\tilde{\beta}_2 \over 3\tilde{\alpha}^2} \,
          \left( \half(\tilde{M}_{K^0}^2+\tilde{M}_{K^+}^2)
                             -\tilde{M}_{\pi^+}^2
                 + 2(\tilde{M}_{K^0}^2-\tilde{M}_{K^+}^2)
                 - {3(\tilde{M}_{K^0}^2-\tilde{M}_{K^+}^2)^2
                    \over 
                       \half(\tilde{M}_{K^0}^2+\tilde{M}_{K^+}^2)
                              -\tilde{M}_{\pi^+}^2 } \right) \,.
                                                             \nonumber
\end{eqnarray}
Again together with $(\tilde{\beta}_1 + 3\tilde{\beta}_2) / 3\tilde{\alpha}^2
\sim -0.01$ and
$\half(\tilde{M}_{K^0}^{*\,2}+\tilde{M}_{K^+}^{*\,2}) -\tilde{M}_{\pi^+}^{*\,2}
\sim 1$, this means that the correction NLO term for $\delta m_+$ is
also small.}
i.e\ $\sim 3\%$ and tends to reduce the value of $\delta m_-$ slightly.
(See next section for the numerical values.)

%----------------------------------------------------------------------------

\section{Physical values of the quark masses}
\label{phy_qm}

%----------------------------------------------------------------------------

To proceed further we now need to substitute eqs.~(\ref{iter_dmpm}) into
eqs.~(\ref{Nucsplit})--(\ref{Xisplit}) to give the pure QCD
contribution to baryon octet  mass splittings in terms of the
pseudoscalar octet masses. 

Before considering this however (see section~\ref{split_qcd}), we shall
first discuss electromagnetic effects and determine the physical values
of the quark masses $\delta m_u^*$ and $\delta m_d^*$ given the
experimental values of the pseudoscalar masses. This will enable
us to investigate the convergence of the $SU(3)$ flavour breaking
expansion. The experimental masses are,
\cite{nakamura10b},
\begin{eqnarray}
   M_{\pi^+}^{\exp}&=& 0.13957018(35)\,\mbox{GeV}\,,
                                                           \nonumber \\
   M_{K^0}^{\exp}  &=& 0.497672(31)\,\mbox{GeV}\,,
                                                           \nonumber \\
   M_{K^+}^{\exp}  &=& 0.493677(16)\,\mbox{GeV}\,,
\label{ps_expt}
\end{eqnarray}
(with as already mentioned $M_{\overline{K}^0} = M_{K^0}$, $M_{K^-} = M_{K^+}$
and $M_{\pi^-} = M_{\pi^+}$), on the outer octet ring, and at the centre
\begin{eqnarray}
   M_{\pi^0}^{\exp} &=& 0.1349766(6)\, \mbox{GeV}\,.
\end{eqnarray}   

We now need to consider electromagnetic effects (which may now
be comparable to the $u$ -- $d$ quark mass difference which is also small).
Electromagnetic effects tend to increase the mass of charged particles
(due to the photon cloud). As a help to estimate these unknown effects,
we use Dashen's theorem, \cite{dashen69a}, which states that if
electromagnetic effects are the only source of breaking of isospin symmetry
(i.e.\ $m_u = m_d$), the leading  electromagnetic energy contribution
to the neutral pseudoscalar particles, i.e.\ the $\pi^0$, $K^0$, vanishes,
while that due to the charged particles, i.e.\ the $\pi^+$, $K^+$ is equal.
As the mass difference in $\pi^0$ and $\pi^+$ due to the $u$ -- $d$ quark
mass difference is negligible, $O(0.1\,\mbox{MeV})$, e.g.\ \cite{gasser82a},
we can thus write 
(e.g.\ \cite{cheng84a,leutwyler00a})
\begin{eqnarray}
   M_{\pi^+}^{\exp\,2}
        &=& M_{\pi^+}^{*\,2} + \mu_\gamma\,, \qquad
               M_{\pi^0}^{\exp\,2} = M_{\pi^0}^{*\,2} \equiv M_{\pi^+}^{*\,2} \,,
                                                           \nonumber \\
   M_{K^+}^{\exp\,2} 
        &=& M_{K^+}^{*\,2} + \mu_\gamma\,, \qquad
               M_{K^0}^{\exp\,2} = M_{K^0}^{*\,2} \,.
\label{dashen_rel}
\end{eqnarray}
(The $^*$ denotes values at the physical point for the pure QCD case.)
Dashen's theorem has corrections of $O(\alpha_{\qed}m_q)$ from higher
order terms. Sometimes possible violations of the theorem are
parametrised by \cite{divitiis11a,colangelo10a}
\begin{eqnarray}
   M_{K^0}^{*\,2} - M_{K^+}^{*\,2} 
      = \left(M_{K^0}^2 - M_{K^+}^2\right)^{\exp}
        + (1 + \epsilon_\gamma)\left( M_{\pi^+}^2
             - M_{\pi^0}^2 \right)^{\exp} \,,
\label{dashen_violation}
\end{eqnarray}
where $\epsilon_\gamma = 0$ corresponds to Dashen's theorem. For example
a positive value for $\epsilon_\gamma$ would thus tend to increase
slightly the value of $M_{K^0}^{*\,2} - M_{K^+}^{*\,2}$. From the first
equation in eq.~(\ref{iter_dmpm}) this would only affect the leading
term in our expansion, so our main result in section~\ref{phys_res}
will be given in terms of the kaon mass splitting,
$M_{K^0}^{*\,2} - M_{K^+}^{*\,2}$ and where we shall consider $\epsilon_\gamma$
as an additional error.

Inverting the relations in eq.~(\ref{dashen_rel}) gives
\cite{nakamura10b}
\begin{eqnarray}
   M_{\pi^+}^{*\,2} = M_{\pi^0}^{\exp\,2}\,, \quad
   M_{K^+}^{*\,2}  = M_{K^+}^{\exp\,2} - (M_{\pi^+}^{\exp\,2}-M_{\pi^0}^{\exp\,2})\,,\quad
   M_{K^0}^{*\,2}  = M_{K^0}^{\exp\,2} \,,
\label{kaon_pureqcd}
\end{eqnarray}
or     
\begin{eqnarray}
   M_{\pi^+}^* = 0.13498 \,\mbox{GeV}\,, \quad
   M_{K^+}^*  = 0.49240 \, \mbox{GeV}\,, \quad
   M_{K^0}^*  = 0.49767 \, \mbox{GeV} \,,
\label{ps_masses_pureqcd}
\end{eqnarray}
which we shall use as the pure QCD values due to differences in the
quark masses with the electromagnetic effects subtracted away
(assuming Dashen's theorem). This gives from eq.~(\ref{Xpi_def})%
\footnote{In \cite{bietenholz11a} we used the average kaon and
pion masses as we were strictly in the $2+1$ flavour case. Here
we need to take into account the differences between charged
and neutral mesons.},
\begin{eqnarray}
   X_\pi^* = 0.4116\,\mbox{GeV}\,,
\label{Xpi_phys}
\end{eqnarray}
(of course this is very close to the experimental value of
$X_\pi^{\exp} = 0.4126\,\mbox{GeV}$).

In Table~\ref{qm_physical} using the masses given in
eq.~(\ref{ps_masses_pureqcd})
\begin{table}[htb]
   \begin{center}
      \begin{tabular}{c|cc|ccc}
              & $\delta m_d^* - \delta m_u^*$ & $\delta m_d^* + \delta m_u^*$
              & $\delta m_u^*$ & $\delta m_d^*$ & $\delta m_s^*$    \\
         \hline
 LO      & 0.0007485(14) & -0.02168(4) & -0.01121(2) & -0.01047(2)
                                                     & 0.02168(4) \\
 NLO$^a$ & 0.0007245(27) & -0.02204(6) & -0.01138(3) & -0.01066(3)
                                                     & 0.02204(6)  \\
 \it{NLO$^b$}
         &\it{0.0007249} &\it{-0.02204} &\it{-0.01138}
                             &\it{-0.01066}  &\it{0.02204}  \\
      \end{tabular}
   \end{center}
\caption{Results for the bare quark mass in lattice units at the
         physical point using eq.~(\protect\ref{ps_masses_pureqcd})
         as input. The first line is the LO result. The next line,
         NLO$^a$, uses the result of eq.~(\protect\ref{iter_dmpm}),
         while as a check the last line in italics, NLO$^b$,
         iterates eqs.~(\protect\ref{K0-Kstar_phys}) and
         (\protect\ref{Kstar-pistar_phys}).}
\label{qm_physical}
\end{table}
we give our results, first giving the LO results for the quark masses,
then the next line, NLO$^a$, gives the results from eq.~(\ref{iter_dmpm}),
which we will be using here. As a check, the third line compares this
NLO result to the NLO result using the Newton--Raphson method to iterate
eqs.~(\ref{K0-Kstar_phys}) and (\ref{Kstar-pistar_phys}).
We find the Newton--Raphson procedure converges very quickly (after one
or two iterations) and find good agreement between the two results
(well within the error bars of NLO$^a$ in Table~\ref{qm_physical})
and so we can be confident that eq.~(\ref{iter_dmpm}) is a good
approximation for the inversion of eqs.~(\ref{K0-Kstar_phys}) and
(\ref{Kstar-pistar_phys}).

We also note that the differences between the LO and NLO are small
of the order of a few percent, indicating that the $SU(3)$ flavour
expansion about the flavour symmetric point appears to be highly convergent.
(We shall discuss this a little further in section~\ref{comparison_fan}).)

%----------------------------------------------------------------------------

\section{Comparison with `fan' plots}
\label{comparison_fan}

%----------------------------------------------------------------------------

We now compare the fit results with the mass values from previous
`fan' plots, \cite{bietenholz11a}, which describe the evolution
of the pseudoscalar and baryon octet masses along a path
from the $SU(3)$ symmetric point down to the physical point.
In \cite{bietenholz11a} the isospin degenerate limit,
i.e.\ $m_u = m_d \equiv m_l$, was considered. So for this
comparison we take the physical quark mass, in lattice units, as
\begin{eqnarray}
   \delta m_l^* \equiv \half(\delta m_u^* + \delta m_d^*)
                = -0.01102(3) \,.
\label{delta_ml_star}
\end{eqnarray}

\begin{figure}[htb]
   \begin{center}
      \includegraphics[width=9.00cm]
         {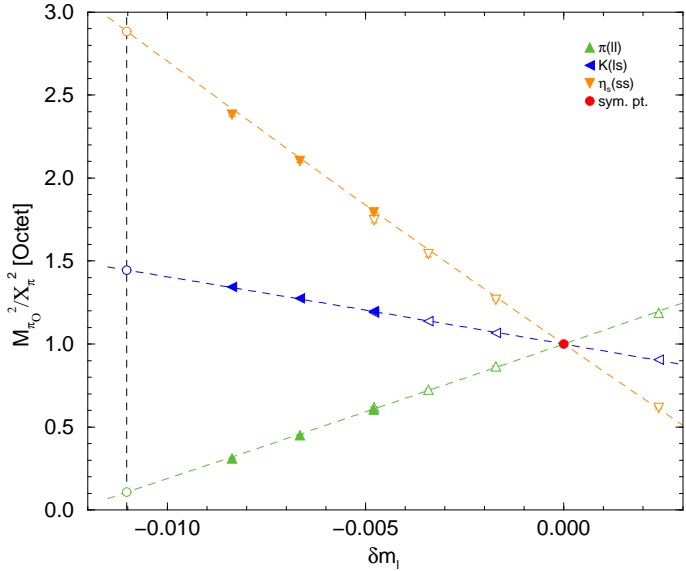}
   \end{center}
\caption{The pseudoscalar meson octet `fan' plot,
         $M_{\pi_O}^2/X_\pi^2$ ($\pi_O = \pi$, $K$, $\eta_s$) versus
         $\delta m_l$. The numerical mass values are taken from
         \protect\cite{bietenholz11a} where filled up triangles,
         left triangles, down triangles are $\pi(ll)$, $K(ls)$,
         $\eta_s(ss)$ results respectively using $32^3\times 64$
         sized lattices. The common symmetric point is the filled
         circle. The opaque up triangles, left triangles,
         down triangles are from $24^3\times 48$ sized lattices
         (and not used in the fits here). The quadratic fits are
         taken from eq.~(\protect\ref{pq_fit_meson_2+1}), together
         with Table~\protect\ref{alpha_beta_results}. The
         vertical line from eq.~(\protect\ref{delta_ml_star}) is the
         pure $N_f = 2+1$ QCD physical point, while the opaque circles
         are the pure QCD hadron mass ratios for $2+1$ quark flavours.}
\label{b5p50_dml_mpsO2oXpi2-jnt_isobrk}
\end{figure}

In Fig.~\ref{b5p50_dml_mpsO2oXpi2-jnt_isobrk} we
show numerical results between the $SU(3)$ flavour symmetric point and
the `physical point' for the numerical pseudoscalar octet on the unitary line
(keeping $\overline{m} = \mbox{const.}$) from \cite{bietenholz11a}.
These masses are compared to the quadratic fit using
eq.~(\ref{pq_fit_meson_2+1}) (i.e.\ together with just the results
of Table~\ref{alpha_beta_results}) for the $2+1$ flavour case,
i.e.\ $m_u = m_d = m_l$. The NNLO terms in the $SU(3)$ flavour
symmetric expansion can be safely ignored in the small
$\delta m_l$ range. Compare the scale of the $x$-axis of
Fig.~\ref{dmua_mNaaaoXN} with that of
Fig.~\ref{b5p50_dml_mpsO2oXpi2-jnt_isobrk}.
We consider $M_\pi(l\overline{l})$, $M_K(l\overline{s})$
and the fictitious PQ particle $\eta_s$, with mass
$M(s\overline{s})$. The comparison is satisfactory.
We also show results from \cite{bietenholz11a} on smaller
$24^3\times 48$ sized lattices. This shows that finite size effects
for these ratios are rather small. Of course, the value of
$\delta m_l^*$ just serves to define pseudoscalar meson mass ratios at
the `physical point' in a $N_f = 2+1$ flavour world. For completeness
we give these numbers here: $\tilde{M}_\pi^{*\,2} = 0.1077(26)$,
$\tilde{M}_K^{*\,2} = 1.446(1)$ and $\tilde{M}_{\eta_s}^{*\,2} = 2.884(4)$.

\begin{figure}[htb]
   \begin{center}
      \includegraphics[width=9.00cm]
         {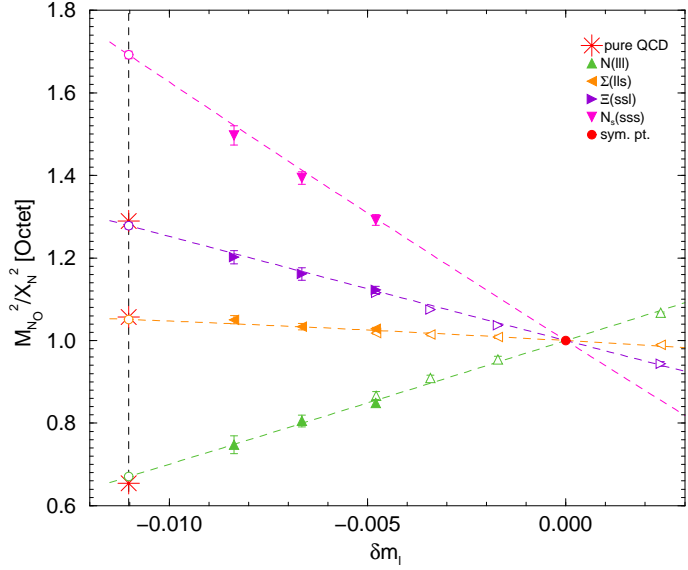}
   \end{center}
\caption{The baryon octet `fan' plot,
         $M_{N_O}^2/X_N^2$ ($N_O = N$, $\Sigma$, $\Xi$, $N_s$) versus
         $\delta m_l$. The numerical mass values are taken from
         \protect\cite{bietenholz11a} where filled up-triangles,
         left-triangles, right-triangles, down-triangles are
         $N(lll)$, $\Sigma(lls)$, $\Xi(ssl)$, $N_s(sss)$ results
         respectively using $32^3\times 64$ sized lattices.
         The common symmetric point is the filled
         circle. The opaque up-triangles, left-triangles,
         right-triangles, down-triangles are from $24^3\times 48$
         sized lattices (and not used in the fits here).
         The quadratic fits are from eq.~(\protect\ref{N_2+1_pq_expt}),
         together with Table~\protect\ref{Noctet_parms}. The
         vertical line from eq.~(\ref{delta_ml_star}) is the
         $N_f = 2+1$ pure QCD physical point, with the opaque circles being
         the determined pure QCD hadron mass ratios for $2+1$
         quark flavours. For comparison, the stars represent
         the average of the squared masses of the appropriate
         particle on the outer ring of the baryon octet,
         Fig.~\protect\ref{baryon_j=half_octet}, i.e.\
         $M_N^{*\,2}(lll) = (M_n^{\exp\,2}(ddu) + M_p^{\exp\,2}(uud))/2$,
         $M_\Sigma^{*\,2}(lls)
                 = (M_{\Sigma^-}^{\exp\,2}(dds) + M_{\Sigma^+}^{\exp\,2}(uus))/2$,
         $M_\Xi^{*\,2}(ssl) 
                 = (M_{\Xi^-}^{\exp\,2}(ssd) + M_{\Xi^0}^{\exp\,2}(ssu))/2$.}
\label{b5p50_dml_mNOomNOpmSigOpmXiOo3-jnt_bk_isobrk}
\end{figure}

In Fig.~\ref{b5p50_dml_mNOomNOpmSigOpmXiOo3-jnt_bk_isobrk} we
show the comparable baryon octet results.
As well as considering the nucleon mass, $M_N(lll)$, we also
show $M_{\Sigma}(lls)$, $M_{\Xi}(lls)$ and also a fictitious PQ
particle -- $M_{N_s}(sss)$. Again the comparison of the NLO (quadratic) fit,
using the expansion coefficient values given in Table~\ref{Noctet_parms},
to the numerical data is satisfactory. For completeness we give here
the values at the $2+1$ QCD physical point (opaque circles) of
$\tilde{M}^{*\,2}_N = 0.6704(46)$, $\tilde{M}^{*\,2}_\Sigma = 1.051(6)$,
$\tilde{M}^{*\,2}_\Xi = 1.278(8)$ and $\tilde{M}^{*\,2}_{N_s} =1.692(9)$.
For a comparison to these values, the stars in
Fig.~\ref{b5p50_dml_mNOomNOpmSigOpmXiOo3-jnt_bk_isobrk}
represent the average of the squared masses of the appropriate
particle, as defined in the figure caption.

%----------------------------------------------------------------------------

\section{Results and discussion}
\label{results}

%----------------------------------------------------------------------------

We are now in a position to numerically determine the baryon octet mass
splittings due to pure QCD $u$ -- $d$ quark mass differences (see 
section~\ref{split_qcd}, our main result), and then in
section~\ref{phys_res} estimate physical values for the splittings.

%----------------------------------------------------------------------------

\subsection{Baryon octet mass splittings}
\label{split_qcd}

%----------------------------------------------------------------------------

After re-writing quark masses in terms of the pseudoscalar octet masses,
section~\ref{ps_meson_isobrk}, and finding that the expansion is highly
convergent in the relevant quark mass range, e.g.\ Table~\ref{qm_physical},
we now insert the expansion of eq.~(\ref{iter_dmpm}) into
eqs.~(\ref{Nucsplit})--(\ref{Xisplit}) which gives the reasonably
compact results
\begin{eqnarray}
   \tilde{M}_N - \tilde{M}_{N^\prime}
      = \tilde{A}_{N-N^\prime}
        \left[ 1 + \tilde{B}_{N-N^\prime}
                   \left( \half(\tilde{M}_{K^0}^2+\tilde{M}_{K^+}^2)
                                 -\tilde{M}_{\pi^+}^2 \right)
        \right]\left( \tilde{M}_{K^0}^2-\tilde{M}_{K^+}^2 \right) \,,
\label{general_split}
\end{eqnarray}
for the pairs $(N, N^\prime) = (n, p)$, $(\Sigma^-, \Sigma^+)$ and
$(\Xi^-, \Xi^0)$, where
\begin{eqnarray}
   \tilde{A}_{n-p}
      =  {\tilde{A}_1^\prime - 2\tilde{A}_2^\prime \over \tilde{\alpha}} \,,
   \qquad
   \tilde{B}_{n-p} 
      = {2 \over 3\tilde{\alpha}}
          \left( {\tilde{\beta}_1 + 3\tilde{\beta}_2 \over \tilde{\alpha}}
                  - {\tilde{B}_1^\prime - 2\tilde{B}_2^\prime 
                      \over \tilde{A}_1^\prime -2\tilde{A}_2^\prime}
          \right) \,,
\label{n-p_split}
\end{eqnarray}
together with
\begin{eqnarray}
   \tilde{A}_{\Sigma^--\Sigma^+}
      =  {2\tilde{A}_1^\prime - \tilde{A}_2^\prime \over \tilde{\alpha}} \,,
   \qquad
   \tilde{B}_{\Sigma^--\Sigma^+} 
      = {2 \over 3\tilde{\alpha}}
           \left( {\tilde{\beta}_1 + 3\tilde{\beta}_2 \over \tilde{\alpha}}
                    - {2\tilde{B}_1^\prime - \tilde{B}_2^\prime 
                    + 3\tilde{B}_3^\prime
                        \over 2\tilde{A}_1^\prime - \tilde{A}_2^\prime}
            \right) \,,
\label{sig-sig_split}
\end{eqnarray}
and
\begin{eqnarray}
   \tilde{A}_{\Xi^--\Xi^0}
      =  {\tilde{A}_1^\prime + \tilde{A}_2^\prime \over \tilde{\alpha}} \,,
   \qquad
   \tilde{B}_{\Xi^--\Xi^0} 
      = {2 \over 3\tilde{\alpha}}
           \left( {\tilde{\beta}_1 + 3\tilde{\beta}_2 \over \tilde{\alpha}}
                    - {\tilde{B}_1^\prime + \tilde{B}_2^\prime 
                    + 3\tilde{B}_3^\prime
                        \over \tilde{A}_1^\prime + \tilde{A}_2^\prime}
            \right) \,.
\label{xi-xi_split}
\end{eqnarray}

As discussed before, as we have unknown electromagnetic effects,
then we shall first present our results in a general form with (known)
coefficients between the baryon and pseudoscalar meson splittings,
within a pure QCD context. We now insert the numerically determined
values from Tables~\ref{Noctet_parms} and \ref{alpha_beta_results} into
eq.~(\ref{general_split}), together with
eqs.~(\ref{n-p_split})--(\ref{xi-xi_split}), to give
\begin{eqnarray}
   \tilde{M}_n - \tilde{M}_p
      &=& 0.0789(41)(8)(8)(32) \, \left( \tilde{M}_{K^0}^2-\tilde{M}_{K^+}^2
                          \right)
                                                           \nonumber \\
      & & \hspace*{0.40in} \times
            \left[ 1 +0.0817(92)
                   \left( \half(\tilde{M}_{K^0}^2+\tilde{M}_{K^+}^2)
                                 -\tilde{M}_{\pi^+}^2 \right)
            \right] \,,
\end{eqnarray}
together with
\begin{eqnarray}
   \tilde{M}_{\Sigma^-} - \tilde{M}_{\Sigma^+}
      &=& 0.2243(35)(22)(2)(90) \, \left( \tilde{M}_{K^0}^2-\tilde{M}_{K^+}^2
                        \right)
                                                           \nonumber \\
      & & \hspace*{0.40in} \times
            \left[ 1 + 0.0077(30)
                   \left( \half(\tilde{M}_{K^0}^2+\tilde{M}_{K^+}^2)
                                 -\tilde{M}_{\pi^+}^2 \right)
            \right] \,,
\end{eqnarray}
and
\begin{eqnarray}
   \tilde{M}_{\Xi^-} - \tilde{M}_{\Xi^0}
      &=& 0.1455(24)(13)(6)(58) \, \left( \tilde{M}_{K^0}^2-\tilde{M}_{K^+}^2
                        \right)
                                                           \nonumber \\
      & & \hspace*{0.40in} \times
            \left[ 1 -0.0324(50)
                   \left( \half(\tilde{M}_{K^0}^2+\tilde{M}_{K^+}^2)
                                 -\tilde{M}_{\pi^+}^2 \right)
            \right] \,,
\end{eqnarray}
as our main numerical result. This is a pure QCD result: the isospin
breaking is just due to the different masses of the $u$, $d$ and $s$ quarks.
(To recapitulate, $^*$ means at the physical point and for the baryon
octet $\tilde{M}_n = M_n/X_N$, etc.\ and for the pseudoscalar octet
$\tilde{M}_{K^0} = M_{K^0}/X_\pi$, etc.\ where $X^2$ is the `average' hadron
$(\mbox{mass})^2$ of the `outer' ring of the octet, given numerically in
eqs.~(\ref{XN_phys}) and (\ref{Xpi_phys}).) Even with the long lever arm of
PQ some of the NLO terms, i.e.\ for $\Sigma^- - \Sigma^+$ are only
marginally determined.

The first error is statistical; the other errors are systematic
as discussed in Appendix~\ref{systematic}. The second error bar
is due to possible finite size effects, the third error estimates
a possible error due to convergence of the $SU(3)$ flavour breaking
expansion, while the last error is due to the choice of path to
the physical point.

The above results are valid for a range of quark masses; we
shall now specialise to the physical point, as discussed
in section~\ref{phy_qm}.

%----------------------------------------------------------------------------

\subsection{Physical values of the mass splittings}
\label{phys_res}

%----------------------------------------------------------------------------

We first note that the NLO term is always small of the order of a
few percent, and only slowly decreases for increasing number of
strange quarks in the baryon. Using for the NLO term the mass values
given in eq.~(\ref{ps_masses_pureqcd}) for the pure QCD case we have
$\half(\tilde{M}_{K^0}^{*\,2}+\tilde{M}_{K^+}^{*\,2})-\tilde{M}_{\pi^+}^{*\,2}
= 1.339$ (using the experimental rather than pure QCD masses
would introduce negligible additional errors into the NLO term).
In addition, using $X_N^{*\,2}$ and $X_\pi^{*\,2}$ from eqs.~(\ref{XN_phys})
and (\ref{Xpi_phys}), this gives
\begin{eqnarray}
   M_n^* - M_p^*
      &=& 0.599(32)(34) \, ( M_{K^0}^{*\,2}-M_{K^+}^{*\,2} ) \,,
                                                           \nonumber \\
   M_{\Sigma^-}^* - M_{\Sigma^+}^*
      &=& 1.553(25)(63) \, ( M_{K^0}^{*\,2}-M_{K^+}^{*\,2} ) \,,
                                                           \nonumber \\
   M_{\Xi^-}^* - M_{\Xi^0}^*
      &=& 0.954(18)(41) \, ( M_{K^0}^{*\,2}-M_{K^+}^{*\,2} ) \,,
\end{eqnarray}
where all masses in these equations are now in $\mbox{GeV}$,
as our final result in terms of the pseudoscalar kaon masses.
The first error is statistical, while the second is the total
systematic error.

Using again the kaon values from eq.~(\ref{ps_masses_pureqcd})
and regarding possible violations of Dashen's theorem,
eq.~(\ref{dashen_violation}), as a further systematic error,
our isospin breaking effects due to pure QCD alone (in $\mbox{MeV}$) are
\begin{eqnarray}
   M_n^* - M_p^* &=& 3.13(15)(16)(76|\epsilon_\gamma|) \,\mbox{MeV} \,,
                                                           \nonumber \\
   M_{\Sigma^-}^* - M_{\Sigma^+}^*
                &=& 8.10(14)(33)(193|\epsilon_\gamma|) \, \mbox{MeV} \,,
                                                           \nonumber \\
   M_{\Xi^-}^* - M_{\Xi^0}^*
                &=& 4.98(10)(21)(120|\epsilon_\gamma|) \, \mbox{MeV} \,.
\label{QCD_md}
\end{eqnarray}
In general, comparing eq.~(\ref{QCD_md}) with eq.~(\ref{expt_mass_split})
would indicate that electromagnetic effects for $n-p$ are negative,
for $\Sigma^--\Sigma^+$ are small and for $\Xi^--\Xi^0$ are positive.

The uncertainty due to $\epsilon_\gamma$ is the dominant error once
we convert to $\mbox{MeV}$ (i.e.\ unknown EM effects are the largest
source of error). As an example, taking $\epsilon_\gamma = 0.7$,
\cite{divitiis11a,colangelo10a} gives
$M_n^* - M_p^* = 3.13(15)(16)(53)\,\mbox{MeV}$,
$M_{\Sigma^-}^* - M_{\Sigma^+}^* = 8.10(14)(33)(135)\,\mbox{MeV}$
and $M_{\Xi^-}^* - M_{\Xi^0}^* = 4.98(10)(21)(84)\,\mbox{MeV}$.

In \cite{walker-lourd12a}, a determination of the $n - p$
isospin breaking effects due to electromagnetic effects was given, with a
result of $-1.30(47)\,\mbox{MeV}$. Adding this to the result
of eq.~(\ref{QCD_md}) gives
\begin{eqnarray}
   (M_n - M_p)^{* + \qed} = 1.83(52)(76|\epsilon_\gamma|)\,\mbox{MeV} \,.
\label{n-p+QED}
\end{eqnarray}
This is to be compared with the experimental result given in
eq.~(\ref{expt_mass_split}) of $1.29\,\mbox{MeV}$. Thus we find
consistency (within errors even with $|\epsilon_\gamma| \approx 0$).
Thus this result also indicates that violations of Dashen's theorem
seem to be small.

\begin{figure}[htb]
   \begin{center}
      \includegraphics[width=6.50cm]
         {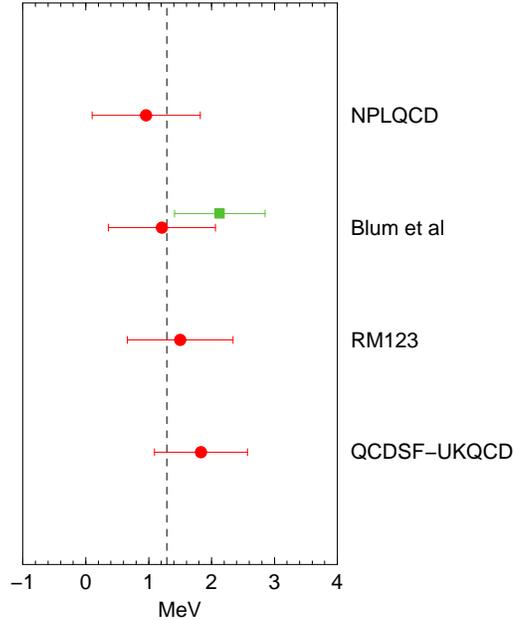}
   \end{center}
\caption{Comparison of the $n$ -- $p$ mass difference of the
         present result (QCDSF-UKQCD or lowest plotted number) with
         NPLQCD, Blum et al., and RM123 
         \protect\cite{beane06a,blum10a,divitiis11a}, respectively
         (top to bottom plotted numbers). The filled circles use
         the QED determination of \cite{walker-lourd12a},
         while the filled square gives the full QCD and QED determination
         from Blum et al.\,. The vertical dashed line is the experimental
         result given in eq.~(\protect\ref{expt_mass_split}).} 
\label{n-p_diff+QED_epg.eq.0p7}
\end{figure}

In Fig.~\ref{n-p_diff+QED_epg.eq.0p7} we compare this $n$ -- $p$ mass
difference including QED, $(M_n - M_p)^{* + \qed}$,
eq.~(\ref{n-p+QED}) with $\epsilon_\gamma = 0.7$,
together the results of \cite{beane06a,blum10a,divitiis11a}.
The filled circles use the QED determination of \cite{walker-lourd12a},
while the filled square includes the full determination from that reference.
Despite the fact that QED effects are treated slightly differently
in each work, good agreement amongst the various determinations and
with the experimental result is found.

%----------------------------------------------------------------------------

\subsection{Conclusions}

%----------------------------------------------------------------------------

In this article we have introduced a method to determine
the isospin breaking effects due to QCD for octet baryons.
Using an $SU(3)$ flavour symmetry breaking expansion in the quark
masses, the pseudoscalar meson octet expansion coefficients
can be determined, which leads to an estimate at the physical
quark mass point. This can then be used together with the
equivalent $SU(3)$ flavour symmetry breaking expansion for the baryon
octet to determine the baryon mass splittings. The expansion
coefficients depend only on the average quark mass $\overline{m}$.
Thus we can use degenerate sea quark masses (either with mass
degenerate $u$ and $d$ quarks or with mass degenerate $u$, $d$ and $s$
quarks, i.e.\ at the flavour symmetric point) provided that
$\overline{m}$ remains unchanged. These are computationally
cheaper simulations than those with mass non-degenerate $u$, $d$
and $s$ quarks.

A further advantage of our procedure is that the mass expansion formulas
are easily generalised to the case of differing valence quark masses
to sea quark masses (i.e.\ partially quenched valence quarks).
This allows an extension of the quark mass range to heavier quark
masses, so that both LO and NLO coefficients (i.e.\ linear terms
and quadratic terms in quark masses) can be determined.
Our final results are given in eq.~(\ref{QCD_md}).
As the NLO turns out to be only a small correction to the LO
this gives us confidence that the $SU(3)$ flavour symmetry
expansion appears to be a highly convergent series.

%----------------------------------------------------------------------------

\section*{Acknowledgements}

%----------------------------------------------------------------------------

The numerical configuration generation was performed using the
BQCD lattice QCD program \cite{nakamura10a} on the IBM
BlueGeneL at EPCC (Edinburgh, UK), the BlueGeneL and P at
NIC (J\"ulich, Germany), the SGI ICE 8200 at
HLRN (Berlin-Hannover, Germany), and the JSCC (Moscow, Russia).
The BlueGene codes were optimised using Bagel \cite{boyle09a}.
The Chroma software library \cite{edwards04a} was used in the data
analysis. This investigation has been supported partly by the DFG
under Contract No.\ SFB/TR 55 (Hadron Physics from Lattice QCD)
and by the EU Grants No.\ 227431 (Hadron Physics2),
and No.\ 238353 (ITN STRONGnet). JN was partially
supported by EU Grant No.\ 228398 (HPC-EUROPA2). JMZ is supported
by the Australian Research Council Grant No.\ FT100100005.
We thank all funding agencies.

%----------------------------------------------------------------------------

\clearpage

\appendix

\section*{Appendix}

%----------------------------------------------------------------------------

\section{Systematic errors}
\label{systematic}

%----------------------------------------------------------------------------

We now consider in this Appendix possible sources of systematic errors:
finite lattice volume, convergence of the $SU(3)$ flavour breaking expansion,
the path to the physical point and finite lattice spacing.

%----------------------------------------------------------------------------

\subsubsection*{Finite lattice volume}

%----------------------------------------------------------------------------

Comparing the available $24^3\times 48$ with the $32^3\times 64$
lattice data in Fig.~\ref{b5p50_dml_mNOomNOpmSigOpmXiOo3-jnt_bk_isobrk}
indicates that for these mass ratios finite size
effects are small. (Finite volume effects were also discussed
in \cite{bietenholz11a} in section 8.3.1.) We now use these unitary
results to estimate possible finite size effects.
For the lightest $24^3\times 48$ point, namely,
$(\kappa_l, \kappa_s) = (0.121040,0.120620)$, $M_\pi L \sim 3.4$
(where $L$ is the spatial lattice size, here $24a$), and the mass
ratio $\tilde{M}$ has finite size error $\sim 1\%$, from comparing the
$24^3\times 48$ lattice result with the $32^3\times 64$ lattice result.
(Either using the results of \cite{bietenholz11a} directly or
equivalently taking the square root of the results in
Fig.~\ref{b5p50_dml_mNOomNOpmSigOpmXiOo3-jnt_bk_isobrk}.)
On the lightest $32^3\times 64$ lattice point, namely $(0.121145,0.120413)$
we have $M_\pi L \sim 3.1$, so we expect the finite size errors
in the mass ratio $\tilde{M}_N$ will be approximately the same,
i.e.\ also $\sim 1\%$. We use this in section~\ref{results}
to estimate systematic errors arising from finite volume effects.

%----------------------------------------------------------------------------

\subsubsection*{$SU(3)$ flavour breaking expansion}

%----------------------------------------------------------------------------

We first note that in Fig.~\ref{b5p50_dml_mNOomNOpmSigOpmXiOo3-jnt_bk_isobrk}
in the range $|\delta m_l| \lsim 0.01$ (and $|\delta m_s| \lsim 0.02$),
there is little curvature. This is in agreement with the $SU(3)$ flavour
breaking expansion, eq.~(\ref{Ntwid2_1+1+1_exp}), where each order
is multiplied by a further $\delta m_q$. From Table~\ref{Noctet_parms}
for the $\tilde{A}_i$ and $\tilde{B}_i$ coefficients (and for the NNLO
order the $\tilde{C}_i$ coefficients) we see there that they remain
approximately all of the same order, so we expect that every increase
in the order leads to a decrease by about an order of magnitude in the
series. This is confirmed in the present
case as we have compared the NNLO determination of $M_N$, $M_\Sigma$
and $M_\Xi$ with the NLO results (removing the heavier quark
mass points until the $\chi^2/\mbox{dof}$ is approximately the same).
The change in $M_N$ was less than half a percent and for
$M_\Sigma$, $M_\Xi$ less, which is equivalent to using
$\sim 10\%$ of the NLO term to estimate systematic errors.

A further example to illustrate this convergence is given by eq.~(42)
of \cite{bietenholz11a}. Here we can form sums and differences
of the decuplet masses which are of order $\delta m_l^0$,
$\delta m_l^1$, $\delta m_l^2$ and $\delta m_l^3$. We see that
each time we add a factor of $\delta m_l$ the quantity decreases
by an order of magnitude (in fact usually by a factor of $\sim 20$),
and the $~O(\delta m_l^3)$ quantity is about $2000$ times smaller
than the leading order quantity. So we believe that convergence
is very good for hyperons. Such an expansion is very good compared
to most approaches available to QCD.

%----------------------------------------------------------------------------

\subsubsection*{Path to physical point}

%----------------------------------------------------------------------------

As also discussed in \cite{bietenholz11a}, the chosen trajectory
in the $m_s$ - $m_l$ plane (keeping $\overline{m}$ constant),
appears to not quite go through the physical point. Using $X_\pi$
(see eq.~(\ref{Xpi_lsdef})) and $X_N$ then from Table~15 of
\cite{bietenholz11a}, we see that $(aX_\pi/aX_N)\times (X_N/X_\pi)^*$
deviates from $1$ by $\sim 4\%$. We use this as an estimate
in section~\ref{results} of systematic errors due to this effect.

%----------------------------------------------------------------------------

\subsubsection*{Finite lattice spacing}

%----------------------------------------------------------------------------

\begin{figure}[htb]
   \begin{center}
      \includegraphics[width=9.00cm]
         {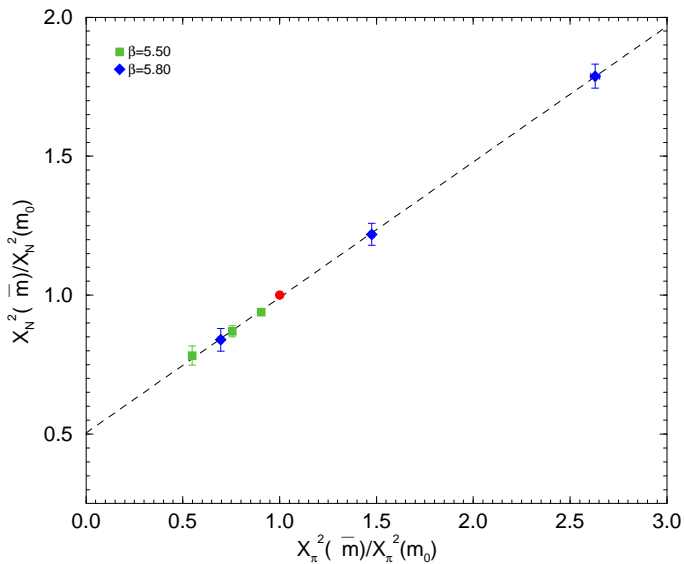}
   \end{center}
\caption{$X_N^2(\overline{m})/X_N^2(m_0)$ against
         $X_\pi^2(\overline{m})/X_\pi^2(m_0)$ along the symmetric
         line. Square symbols are the $\beta = 5.50$ results
         and are given in \protect\cite{bietenholz11a} and
         \protect\cite{horsley11a}, while diamonds are the
         $\beta = 5.80$ results \protect\cite{QCDSF12a}.
         (All results are on $32^3\times 64$ sized lattices.)
         The point where $m_0 = \overline{m}$
         ($\kappa_0 = 0.12090$ for $\beta = 5.50$ and
         $\kappa_0 = 0.12281$ for $\beta = 5.80$)
         is denoted by a circle.}
\label{b5p50+b5p80_mps2omps20_mN2omN20}
\end{figure}

Non-perturbative $O(a)$ improved clover fermions are employed in order
to minimise finite lattice spacing effects in the mass ratios
determined here. Any effects are difficult to estimate if only one
$\beta$ (or $a$ value) is available, so as a check we have performed
some additional simulations at $\beta = 5.80$ (with an estimated
$a \sim 0.06\,\mbox{fm}$). The results are along the $SU(3)$ flavour
symmetric line (as three light quarks might show effects sooner
than two light and one heavier quark). Again we have
located (as described in \cite{bietenholz11a}) the starting point
on this line, $m_0$ for a path in the $m_s$--$m_l$ plane
leading to the physical point and then considered comparable
mass ratios for the nucleon, $X_N^2(\overline{m})/X_N^2(m_0)$ (against
$X_\pi^2(\overline{m})/X_\pi^2(m_0)$), where we denote here a general
point on the $SU(3)$ symmetric line by $\overline{m}$. In
Fig.~\ref{b5p50+b5p80_mps2omps20_mN2omN20} we show these results.
Both $\beta$ values lie on a common line and show no systematic
lattice spacing dependence and so lie close to the continuum
limit (certainly within the precision achievable here).

%----------------------------------------------------------------------------

%\clearpage

%----------------------------------------------------------------------------

\section{Tables}
\label{pq_mass_tables}

%----------------------------------------------------------------------------

We now give in Tables~\ref{pq_N_masses_I} -- \ref{pq_ps_masses} the
baryon and pseudoscalar PQ mass results used in the analysis. Additional
$32^3\times 64$ results along the line $\overline{m}=\mbox{const.}$
also used here are given in Tables~22 and 20 of \protect\cite{bietenholz11a}
for the baryon and pseudoscalar meson particles, respectively.

\begin{table}[htb]
   \begin{center}
      \begin{tabular}{cc|l}
         $\kappa_a$& $\kappa_b$& $M(aab)$ \\ 
         \hline
         0.110000 &  0.110000 &  1.969(3)    \\
         0.110000 &  0.115000 &  1.780(3)    \\
         0.110000 &  0.120000 &  1.555(3)    \\
         0.110000 &  0.120512 &  1.530(3)    \\
         0.110000 &  0.120900 &  1.511(3)    \\
         0.114000 &  0.114000 &  1.520(3)    \\
         0.114000 &  0.116000 &  1.436(3)    \\
         0.114000 &  0.118000 &  1.345(3)    \\
         0.114000 &  0.120000 &  1.245(3)    \\
         0.114000 &  0.120900 &  1.199(3)    \\
         0.115000 &  0.110000 &  1.592(3)    \\
         0.115000 &  0.115000 &  1.397(3)    \\
         0.115000 &  0.120000 &  1.161(2)    \\
         0.115000 &  0.120512 &  1.133(2)    \\
         0.115000 &  0.120900 &  1.113(3)    \\
         0.116000 &  0.114000 &  1.354(3)    \\
         0.116000 &  0.116000 &  1.268(3)    \\
         0.116000 &  0.118000 &  1.175(3)    \\
         0.116000 &  0.120000 &  1.072(3)    \\
         0.116000 &  0.120900 &  1.023(2)    \\
         0.118000 &  0.114000 &  1.173(3)    \\
         0.118000 &  0.116000 &  1.085(3)    \\
         0.118000 &  0.118000 &  0.9887(25)  \\
         0.118000 &  0.120000 &  0.8800(23)  \\
         0.118000 &  0.120900 &  0.8267(23)  \\
      \end{tabular}
   \end{center}
\caption{Results for the PQ baryon masses, $M(aab)$ in lattice
         units at $\beta = 5.50$ from a $32^3\times 64$ lattice with
         sea quark masses at the symmetric point,
         i.e.\ $\kappa_l = \kappa_s = 0.12090$
         and valence quarks $\kappa_a$, $\kappa_b$.}
\label{pq_N_masses_I}
\end{table}

\begin{table}[htb]
   \begin{center}
      \begin{tabular}{cc|l}
         $\kappa_a$& $\kappa_b$& $M(aab)$ \\ 
         \hline
         0.120000 &  0.110000 &  1.134(3)    \\
         0.120000 &  0.114000 &  0.9726(26)  \\
         0.120000 &  0.115000 &  0.9279(24)  \\
         0.120000 &  0.116000 &  0.8812(24)  \\
         0.120000 &  0.118000 &  0.7789(23)  \\
         0.120000 &  0.120000 &  0.6588(23)  \\
         0.120000 &  0.120512 &  0.6232(24)  \\
         0.120000 &  0.120900 &  0.5945(26)  \\
         0.120512 &  0.110000 &  1.081(3)    \\
         0.120512 &  0.115000 &  0.8722(25)  \\
         0.120512 &  0.120000 &  0.5945(25)  \\
         0.120512 &  0.120512 &  0.5564(27)  \\
         0.120512 &  0.120900 &  0.5247(30)  \\
         0.120900 &  0.110000 &  1.041(4)    \\
         0.120900 &  0.114000 &  0.8755(36)  \\
         0.120900 &  0.115000 &  0.8295(32)  \\
         0.120900 &  0.116000 &  0.7811(34)  \\
         0.120900 &  0.118000 &  0.6739(33)  \\
         0.120900 &  0.120000 &  0.5435(32)  \\
         0.120900 &  0.120512 &  0.5031(35)  \\
         0.120900 &  0.120900 &  0.4673(27)  \\
      \end{tabular}
   \end{center}
\caption*{Table~\protect\ref{pq_N_masses_I} continued.}
\end{table}

\begin{table}[htb]
   \begin{center}
      \begin{tabular}{cc|l}
         $\kappa_l$& $\kappa_s$& $M(sss)$ \\ 
         \hline
         0.121040 &  0.120620 &  0.5265(16) \\
         0.121095 &  0.120512 &  0.5446(16) \\
         0.121145 &  0.120413 &  0.5682(13) \\
      \end{tabular}
   \end{center}
\caption{Additional results to Table~22 of \protect\cite{bietenholz11a}
         for the PQ baryon masses, $M_{N_s} \equiv M(sss)$ in lattice
         units from a $32^3\times 64$ lattice along the line
         $\overline{m}=\mbox{const.}$\,.}
\label{pq_Ns_masses}
\end{table}

\begin{table}[htb]
   \begin{center}
      \begin{tabular}{cc|l}
         $\kappa_a$& $\kappa_b$& $M(a\overline{b})$ \\ 
         \hline
         0.110000 &  0.110000 &  1.2485(3)    \\
         0.110000 &  0.115000 &  1.0583(3)    \\
         0.110000 &  0.120000 &  0.8351(4)    \\
         0.110000 &  0.120900 &  0.7909(10)   \\
         0.110000 &  0.120512 &  0.8100(6)    \\
         0.114000 &  0.114000 &  0.9436(3)    \\
         0.114000 &  0.116000 &  0.8583(3)    \\
         0.114000 &  0.118000 &  0.7664(3)    \\
         0.114000 &  0.120000 &  0.6669(4)    \\
         0.114000 &  0.120900 &  0.6200(7)    \\
         0.115000 &  0.115000 &  0.8593(3)    \\
         0.115000 &  0.120000 &  0.6202(4)    \\
         0.115000 &  0.120900 &  0.5720(6)    \\
         0.115000 &  0.120512 &  0.5929(5)    \\
         0.116000 &  0.116000 &  0.7706(3)    \\
         0.116000 &  0.118000 &  0.6754(3)    \\
         0.116000 &  0.120000 &  0.5710(4)    \\
         0.116000 &  0.120900 &  0.5213(6)    \\
         0.118000 &  0.118000 &  0.5752(3)    \\
         0.118000 &  0.120000 &  0.4628(4)    \\
         0.118000 &  0.120900 &  0.4077(5)    \\
         0.120000 &  0.120000 &  0.3342(4)    \\
         0.120000 &  0.120900 &  0.2646(5)    \\
         0.120000 &  0.120512 &  0.2958(4)    \\
         0.120512 &  0.120900 &  0.2174(5)    \\
         0.120512 &  0.120512 &  0.2534(4)    \\
         0.120900 &  0.120900 &  0.1747(5)    \\
      \end{tabular}
   \end{center}
\caption{Results for the PQ pseudoscalar masses, $M(a\overline{b})$
         in lattice units at $\beta = 5.50$ from a $32^3\times 64$
         lattice with sea quark masses at the symmetric point,
         i.e.\ $\kappa_l = \kappa_s = 0.12090$ and valence quarks
         $\kappa_a$, $\kappa_b$.}
\label{pq_ps_masses}
\end{table}

%----------------------------------------------------------------------------

\clearpage

%----------------------------------------------------------------------------

%----------------------------------------------------------------------------


\begin{thebibliography}{55}

%----------------------------------------------------------------------------

\bibitem{nakamura10b}
   K. Nakamura et al. (Particle Data Group), 
   \emph{J.\ Phys.\ G} {\bf 37} (2010) 075021.

\bibitem{beane06a}
   S.~R. Beane, K. Orginos, and M.~J. Savage,
   \emph{Nucl.\ Phys.\ } {\bf B768} (2007) 38,
   [{\tt arXiv:hep-lat/0605014}].

\bibitem{blum10a}
   T. Blum, T. Doi, M. Hayakawa, T. Izubuchi, S. Uno, N. Yamada,
   and R. Zhou,
   \emph{Phys.\ Rev.\ D} {\bf 82} (2010) 094508,
   [{\tt arXiv:1006.1311[hep-lat]}].

\bibitem{divitiis11a}
   G.~M. de Divitiis, P. Dimopoulos, R. Frezzotti, V. Lubicz, G. Martinelli,
   R. Petronzio, G.~C. Rossi, F. Sanfilippo, S. Simula, N. Tantalo,
   and C. Tarantino,
   [RM123 Collaboration],
   \emph{J.\ High Energy Phys.\ } {\bf 04} (2012) 124,
   [{\tt arXiv:1110.6294[hep-lat]}];
   \emph{PoS(Lattice 2011)} (2011) 291,
   {\tt arXiv:1202.5222[hep-lat]}.

\bibitem{duncan96a}
   A. Duncan, E. Eichten, and H. Thacker,
   \emph{Phys.\ Rev.\ Lett.\ } {\bf 76} (1996) 3894,
   [{\tt arXiv:hep-lat/9602005}].

\bibitem{blum07a}
   T. Blum, T. Doi, M. Hayakawa, T. Izubuchi, and N. Yamada,
   \emph{Phys.\ Rev.\ D} {\bf 76} (2007) 114508,
   [{\tt arXiv:0708.0484[hep-lat]}].

\bibitem{basak08a}
    S. Basak, A.Bazavov, C. Bernard, C. DeTar, W. Freeman, S. Gottlieb,
    U.~M. Heller, J.~E. Hetrick, J. Laiho, L. Levkova, J. Osborn,
    R. Sugar, and D. Toussaint,
    \emph{PoS(LATTICE2008)} (2008) 127,
    {\tt arXiv:0812.4486[hep-lat]}.

\bibitem{portelli12a}
    A. Portelli, S. D\"urr, Z. Fodor, J. Frison, C. Hoelbling, S.~D. Katz,
    S. Krieg, T. Kurth, L. Lellouch, T. Lippert, A. Ramos
    and K.~K. Szab{\'o},
    \emph{PoS(Lattice 2011)} (2011) 136,
    {\tt arXiv:1201.2787[hep-lat]}.

\bibitem{glaessle11a}
    B. Glaessle and G.~S. Bali,
    \emph{PoS(Lattice 2011)} (2011) 282,
    {\tt arXiv:1111.3958[hep-lat]}.

\bibitem{aoki12a}
    S. Aoki, K.-I. Ishikawa, N. Ishizuka, K. Kanaya, Y. Kuramashi,
    Y. Nakamura, Y. Namekawa, M. Okawa, Y. Taniguchi, A. Ukawa,
    N. Ukita and T. Yoshi{\'e},
    [PACS-CS Collaboration]
    \emph{Phys.\ Rev.\ } {\bf D86} (2012) 034507,
    [{\tt arXiv:1205.2961[hep-lat]}];
    N. Ukita,
    \emph{PoS(Lattice 2011)} (2011) 144,
    {\tt arXiv:1111.6380[hep-lat]}.

\bibitem{walker-lourd12a}
   A. Walker-Loud, C.~E. Carlson and G.~A. Miller,
   \emph{Phys.\ Rev.\ Lett.\ } {\bf 108} (2012) 232301,
   [{\tt arXiv:1203.0254[nucl-th]}].

\bibitem{bietenholz10a}
   W. Bietenholz, V. Bornyakov, N. Cundy, M. G\"ockeler, R. Horsley,
   A.~D. Kennedy, W.~G. Lockhart, Y. Nakamura, H. Perlt, D. Pleiter,
   P.~E.~L. Rakow, A. Sch\"afer, G. Schierholz, A. Schiller, H. St\"uben,
   and J.~M. Zanotti,
   \emph{Phys.\ Lett.\ B} {\bf 690} (2010) 436,
   [{\tt arXiv:1003.1114[hep-lat]}].

\bibitem{bietenholz11a}
   W. Bietenholz, V. Bornyakov, M. G\"ockeler, R. Horsley, W.~G. Lockhart,
   Y. Nakamura, H. Perlt, D. Pleiter, P.~E.~L. Rakow, G. Schierholz,
   A. Schiller, T. Streuer, H. St\"uben, F. Winter, and J.~M. Zanotti,
   [QCDSF--UKQCD Collaboration],
   \emph{Phys.\ Rev.\ D} {\bf 84} (2011) 054509,
   [{\tt arXiv:1102.5300[hep-lat]}].

\bibitem{QCDSF12a}
   QCDSF--UKQCD Collaboration, in preparation.

\bibitem{gell-mann62a}
   M. Gell-Mann, 
   \emph{Phys.\ Rev.\ } {\bf 125} (1962) 1067.

\bibitem{okubo62a}
   S. Okubo,
   \emph{Prog.\ Theor.\ Phys.\ } {\bf 27} (1962) 949.

\bibitem{gasser82a}
   J. Gasser and H. Leutwyler,
   \emph{Phys.\ Rep.\ } {\bf 87} (1982) 77.

\bibitem{coleman61a}
   S. Coleman and S.~L. Glashow,
   \emph{Phys.\ Rev.\ Lett.\ } {\bf 6} (1961) 423. 

\bibitem{cundy09a}
   N. Cundy, M. G\"ockeler, R. Horsley, T. Kaltenbrunner, A.~D. Kennedy,
   Y. Nakamura, H. Perlt, D. Pleiter, P.~E.~L. Rakow, A. Sch\"afer,
   G. Schierholz, A. Schiller, H. St\"uben, and J.~M. Zanotti,
   [QCDSF--UKQCD Collaboration],
   \emph{Phys.\ Rev.\ D} {\bf 79} (2009) 094507,
   [{\tt arXiv:0901.3302[hep-lat]}].

\bibitem{dashen69a}
   R. Dashen,
   \emph{Phys.\ Rev.\ } {\bf 183} (1969) 1245.

\bibitem{cheng84a}
   T-P. Cheng and L-F. Li,
   Gauge Theory of Elementary Particle Physics, Oxford University Press (1984).

\bibitem{leutwyler00a}
   H. Leutwyler, 
   \emph{Nucl.\ Phys.\ Proc.\ Suppl.\ } {\bf 94} (2001) 108,
   {\tt arXiv:hep-ph/0011049}.

\bibitem{colangelo10a}
   G. Colangelo, S. D\"urr, A. J\"uttner, L. Lellouch, H. Leutwyler,
   V. Lubicz, S. Necco, C.~T. Sachrajda, S. Simula, A. Vladikas,
   U. Wenger, and H. Wittig,
   [FLAG (Flavianet Lattice Averaging Group)],
   \emph{Eur.\ Phys.\ J.\ C} {\bf 71} (2011) 1695,
   [{\tt arXiv:1011.4408[hep-lat]}].

\bibitem{horsley11a}
    R. Horsley, Y. Nakamura, H. Perlt, D. Pleiter, P.E.L. Rakow,
    G. Schierholz, A. Schiller, H. St\"uben, F. Winter, J.~M. Zanotti
    [QCDSF--UKQCD Collaboration],
    \emph{Phys.\ Rev.\ D} {\bf 85} (2012) 034506,
    [{\tt arXiv:1110.4971[hep-lat]}].

\bibitem{nakamura10a}
   Y. Nakamura and H. St\"uben,
   \emph{PoS(Lattice 2010)} (2010) 040,
   {\tt arXiv:1011.0199[hep-lat]}.

\bibitem{boyle09a}
   P.~A. Boyle,
   \emph{Comp.\ Phys.\ Comm.\ } {\bf 180} (2009) 2739.

\bibitem{edwards04a}
   R.~G. Edwards and B. Jo{\'o},
   \emph{Nucl.\ Phys.\ Proc.\ Suppl.\ } {\bf 140} (2005) 832,
   {\tt arXiv:hep-lat/0409003}.

\end{thebibliography}
\end{document}